\documentclass[aps,superscriptaddress,english,12pt,nofootinbib]{revtex4}

\usepackage{float}

\DeclareUnicodeCharacter{2212}{-}

\usepackage{amssymb, amsmath, bm, dcolumn, epsf, graphicx, latexsym, slashed, simplewick}
\usepackage[utf8,utf8x]{inputenc}
\usepackage[normalem]{ulem}
\usepackage{color}

\def\be{\begin{equation}}
\def\ee{\end{equation}}
\def\bea{\begin{eqnarray}}
\def\eea{\end{eqnarray}}

\usepackage{hyperref}
\usepackage{comment}

\newcommand\apjl{{\rm ApJL}}     
\newcommand\apjs{{\rm ApJS}}
\renewcommand\ao{{\rm ApOpt}}
\newcommand\aap{{\rm A\&A}}

\usepackage[normalem]{ulem}

\begin{document}

\widetext


\title{Emergent Unparticles Dark Energy can restore cosmological concordance}

\author{Ido Ben-Dayan}
\author{Utkarsh Kumar}
\affiliation{Physics Department, Ariel University, Ariel 40700, Israel}
\date{\today}

\begin{abstract}
Addressing the discrepancy between the late and early time measurements of the Hubble parameter, $H_0$, and the so-called $S_8$ parameter has been a challenge in precision cosmology. Several models are present to address these tensions, but very few of them can do so simultaneously. In the past, we have suggested Banks-Zaks/Unparticles as an emergent Dark Energy model and claimed that it can ameliorate the Hubble tension. In this work, we test this claim and perform a likelihood analysis of the model and its parameters are given current data and compare it to $\Lambda$CDM. The model offers a possible resolution of Hubble tension and softens the Large Scale Structure (LSS) tension without employing a scalar field or modifying the gravitational sector. Our analysis shows a higher value of $H_0 \sim 70 - 73$ km/sec/Mpc and a slightly lower value of $S_8$ for various combinations of data sets. Consideration of Planck CMB data combined with the Pantheon sample and 
 SH0ES priors lowers the $H_0$ and $S_8$ tension to $0.96  \sigma$ and $0.94 \sigma$ respectively with best-fit $\Delta \chi^2 \approx -10$ restoring cosmological concordance. Significant improvement in the likelihood persists for other combinations of data sets as well. Evidence for the model is given by inferring one of its parameters to be $x_0\simeq-4.36$.

\end{abstract}

\maketitle

\section{Introduction}

The Concordance Model of Cosmology also known as $\Lambda$CDM is well established considering recent cosmological observations \cite{Planck:2018vyg,Planck:2019nip,Planck:2018lbu,Ross:2014qpa,Beutler:2011hx,BOSS:2016wmc,Bautista:2020ahg,Gil-Marin:2020bct,eBOSS:2020yzd,Neveux:2020voa,Hou:2020rse,duMasdesBourboux:2020pck,DES:2021wwk,Pan-STARRS1:2017jku,Riess:2021jrx,Freedman:2019jwv}. The model attributes about $70\%$ of its energy density to a cosmological constant (CC) or more generally Dark Energy (DE). DE drives the present acceleration of the Universe and is a parameterization of our ignorance. Many dynamical models that explain the observed acceleration have been put forward \cite{Copeland:2006wr,Bamba:2012cp,yoo2012theoretical,Zlatev:1998tr}. Nevertheless, the most economical solution - the CC, is still an excellent fit for the data. In the past decade, accumulated evidence from various astrophysical probes raises questions on the validity of $\Lambda$CDM. While the model is consistent with each probe separately, the best-fit value of cosmological parameters defers by several standard deviations. The reasons may be some unaccounted systematic errors in the different measurements \cite{Ben-Dayan:2014swa}, but it may very well be a signal for New Physics \cite{Mortsell:2018mfj,Verde:2019ivm,Freedman:2021ahq,DiValentino:2021izs,Kamionkowski:2022pkx,Agrawal:2019lmo,Lin:2019qug,Smith:2019ihp}. The tensions we are most interested in are the Hubble tension - which is the most pronounced one, at the level of $>4\sigma$ and the $S_8$ tension \cite{DiValentino:2018gcu,Garcia-Garcia:2021unp,LSSTDarkEnergyScience:2022amt,Heymans:2020gsg,DES:2022urg}, at the level of $2-3\sigma$. 
Many attempts to reconcile these discrepancies have been put forward \cite{Alexander:2022own,Knox:2019rjx,Sabla:2021nfy,Sakstein:2019fmf,CarrilloGonzalez:2020oac,Hill:2020osr,Ivanov:2020ril}. However, it seems that many suggestions that reduce the Hubble tension seem to increase the $S_8$ one \cite{DiValentino:2020vvd}. 

Recently \cite{Artymowski:2019cdg,Artymowski:2020zwy,Artymowski:2021fkw}, we have suggested that Dark Energy is due to a Banks-Zaks theory slightly removed from its conformal fixed point and at finite temperature. The theory acts as a perfect fluid. Because the theory is away from the fixed point the energy density and pressure receive a correction that depends on the anomalous dimension and is temperature dependent \cite{Grzadkowski:2008xi}. At high temperatures, the fluid behaves as radiation, and as the temperature decreases with the expansion of the Universe, the correction becomes significant and the fluid behaves as DE and asymptotes to a CC at future infinity. 

It is important to note that this behavior is due to the dynamical behavior of the fluid in the FLRW background. Moreover, the DE behavior is not due to some fundamental degree of freedom, but due to the thermodynamical behavior of the fluid. It is an emergent collective macroscopic phenomenon. As such, it is free of initial conditions and fine-tuning problems that are common in the DE literature and has a built-in tracker mechanism. It does not modify gravity and does not include a scalar field with a potential so it is free of the Swampland conjectures \cite{Artymowski:2019vfy,Ben-Dayan:2018mhe,Garg:2018reu}. Therefore, it is rather unique compared to other DE models in the literature. 
In \cite{Artymowski:2020zwy,Artymowski:2021fkw} we have analyzed the predictions of the model and showed that it is stable. The most notable predictions with respect to $\Lambda$CDM are that there is a contribution to $N_{eff}$ - number of relativistic degrees of freedom at decoupling, and a deviation from $w_{DE}=-1$ - the equation of state of DE, that may be measured in the future. The model also predicts small deviations from the growth rate of perturbations in $\Lambda$CDM, which will be rather difficult to measure. Other interesting constraints can come from its interaction with the CMB \cite{vanPutten:2022may}. Finally, we claimed that the model can ameliorate the Hubble tension.

In this work, we perform a likelihood analysis of our model,  dubbed "Emergent Unparticles Dark Energy" (UDE), and derive constraints on the parameters of the model. We consider several data sets - CMB measurements of Planck alone, as well as adding Planck lensing, BAO, DES, and supernovae data - the Pantheon sample and the SH0ES result.
The model provides a significantly better fit to the data compared to $\Lambda$CDM, at the level of $\Delta \chi^2=-2.1$--$\,-10.4$.
Most importantly, it reduces the Hubble tension with $H_0\gtrsim 70$ km/s/Mpc in accord with supernovae type Ia (SNIa) measurements and slightly reduces the $S_8$ tension to $S_8\simeq 0.78-0.816$. Hence, we have an economical model that relieves existing tensions without introducing new ingredients and is based on collective phenomena. Future cosmological observations can further test the UDE model by measuring $N_{eff}, w_{DE}$ or deviations in the growth of perturbations from the $\Lambda$CDM predictions. 

The manuscript is organized as follows. We first describe in length the existing Hubble and $S_8$ tensions. In section \ref{sec:review} we review the UDE model and its deviation from $\Lambda$CDM. We then list the different data sets that we use in section \ref{sec:data}. In \ref{sec:results} we describe our results. We then conclude in
\ref{sec:conclusions}.

\section{Existing Cosmic Tensions} 
In general, one can consider various tensions or anomalies in cosmological data discussed below. We focus on two celebrated ones - the Hubble tension and the $S_8$ or Large Scale Structure (LSS) tension. 
\subsection{The Hubble ($H_0$) tension} 
The Hubble tension arises from a discrepancy in the measurements of the present value of the Hubble parameter, $H_0$ using probes of the early and late Universe. Assuming $\Lambda$CDM, the \texttt{Planck 2018} CMB data infers $H_0 = 67.4\pm 0.5 $ km/sec/Mpc \cite{Planck:2018vyg}. On the other hand, local direct measurements of SNIa like \texttt{SH0ES} report $H_0 = 73.04 \pm 1.04$ km/sec/Mpc \cite{Riess:2021jrx} for the $\Lambda$CDM Universe. The inferred value of $H_0$ from CMB is derived from the direct measurements of the angular size of the acoustic scale in the CMB power spectrum while \texttt{SH0ES} measurement is the result of the construction of a cosmic ladder (distance-redshift relation). Currently, there is a $\sim 5\,\sigma$ discrepancy between these measurements. However, in addition to the aforementioned experiments, the discrepancy in present-day Hubble value ($H_0$) persists even if some other cosmological probe is used to infer the $H_0$. For example, one can use the combination of data from Big Bang Nucleosynthesis (BBN), Large Scale Structure data, and Baryonic Acoustic Oscillations measurements resulting in a value of $H_0$ that is $3.2 -3.6 \sigma$ away from $\texttt{SH0ES}$  \cite{DES:2017txv,Cuceu:2019for,Schoneberg:2019wmt}.

If one assumes that this tension is not due to any experimental systematics, then the measurements are misinterpreted within the $\Lambda$CDM model. This fact drives the search for New Physics that resolves the tension. The \texttt{SH0ES} measurement is model-independent, which motivates us to build a model which predicts the increase in the CMB-derived Hubble value to the same as \texttt{SH0ES}. 
Tons of efforts have been made to increase the value of $H_0$.  These efforts act in accordance with one of the options to increase the CMB-derived $H_0$ value:  (i) Modification in pre-recombination sound speed, (ii) Modification of the energy density of dark energy before or after recombination, or a combination of both. 

The UDE model and its time-dependent equation of state act as  
a cosmological constant at late times and evolve as radiation at early times, so it belongs to the second type. However, the reduction of the $H_0$ tension is a byproduct of the UDE model, and not an additional tuning or requirement.  We will review the model
in the next section. 
\subsection{The LSS ($S_8$) tension}
In addition to the $H_0$ tension, data from cosmic shear and galaxy clustering surveys which independently constrain the amplitude of variance in matter fluctuations is quantified as $S_8$,
\begin{eqnarray}
S_8 = \sigma_8\,\left(\frac{\Omega_{m}}{0.3}\right)^{0.5}.
\end{eqnarray}
Here $\sigma_8$ is the measure of the rms amplitude of linear matter density fluctuations over a sphere of radius $R = 8 Mpc/h$ today, and $\Omega_{m}$ is the relative matter-energy density of the Universe today. $\sigma_8$ is defined by the following integral:
\begin{eqnarray}
\left< \sigma_8\right> ^{2} = \frac{1}{2\,\pi^{2}} \, \int \frac{d k}{k} \, W^{2} (k R)\,k^3\,P(k)\,,
\end{eqnarray}
where is $P(k)$ being the  linear matter power spectrum calculated today and $W(k R)$ is spherical top-hat filter of radius $R = 8 $ Mpc/h. 

There is $2-3\sigma$ tension between $S_8$ measured from the data from large-scale structure and Planck CMB data. In particular, CMB derived $S_8$ comes out to be $0.834\pm 0.016$ \cite{Planck:2018vyg} while Dark Energy Survey measures $S_8 = 0.776\pm 0.017$ from the combined analysis of the clustering and lensing of foreground and background galaxies respectively \cite{DES:2021wwk}. Weak lensing surveys such as \texttt{KiDS}  report $S_8 = 0.759^{+0.024}_{-0.021}$ \cite{KiDS:2020suj}, and see also \cite{White:2021yvw}
that cross-correlate DESI Luminous Red Galaxies and
Planck CMB lensing resulting in $S_8=0.73\pm 0.03$.

\section{Dark Energy from Unparticles} \label{sec:review}
\subsection{Background Cosmology}  We consider the flat Friedmann-Lemaitre-Robertson-Walker (FLRW) Universe filled with unparticles, radiation, and matter with the Friedmann equations, 
\begin{eqnarray}
3\,H^{2} &=& \rho_r + \rho_m + \rho_u \,, \\
\dot{H} &=& -\frac{1}{2} \left( \frac{4}{3} \rho_r + \rho_m + \rho_u + p_u \right) \,.
\end{eqnarray}
where $\rho_r$ and $\rho_m$ are the energy density of radiation and matter content of the Universe, that scale as $a^{-4}$ and $a^{-3}$ respectively. Unparticles' energy density and pressure are defined as
\begin{eqnarray}
\rho_u &\simeq&  \sigma T^4 +B T^{4+\delta}\equiv\sigma\,T_c^{4}\,y^4\,\left( 1 - \frac{4\,\left( \delta + 3\right)\,y^{\delta}}{3\,\left(\delta + 4\right)}\right), \\
p_u &\simeq&\frac{1}{3}\sigma T^4 +\frac{B}{\delta+3}\, T^{4+\delta} \equiv  \frac{1}{3}\,\sigma\, T_c^{4} \, y^{4}\,\left( 1 - \frac{4\,y^{\delta}}{\delta + 4}\right)\,,
\end{eqnarray}
where $\sigma$ is the "Stefan-Boltzman constant" that measures the number of degrees of freedom, $\delta = a + \gamma$, $a > 0$ is a constant which determines the $\beta$ function around the IR fixed point, $\gamma$ is an anomalous dimension, $y = \frac{T}{T_c}$ is the dimensionless temperature of unparticles and $T_c$, defined in terms of unparticles parameters as $T_c= \left[\frac{4\left( \delta +3\right)}{3\left(\delta + 4\right)}\,\left(-\frac{\sigma}{B}\right)\right]^{1/\delta}$, is the temperature of unparticles at which $\rho_u + p_u = 0$,  \cite{Artymowski:2019cdg,Artymowski:2020zwy}. A unique feature of the model is that the unparticles equation of state is time or temperature dependent. At high temperatures, i.e. for $T \gg \Lambda_{\mathcal{U}}$ (where $\Lambda_{\mathcal{U}}$ is a cut-off scale of the theory), Banks-Zaks (BZ) particles are coupled to the standard model with energy density $\rho = \sigma_{BZ}\, T^{4}$ while below the $\Lambda_{\mathcal{U}}$ scale, the BZ sector decouples from SM and the BZ sector is called unparticles.\footnote{Note that the issue of DE from unparticles has already been partially investigated in \cite{Dai:2009mq,Chen:2009ui,Jamil:2011iu}.
 In \cite{Dai:2009mq} the authors considered scalar unparticles with a mass as a function of  scaling dimension of unparticles. In addition to that Unparticles also have been studied in framework of general relativity and  loop quantum cosmology \cite{Chen:2009ui,Jamil:2011iu} where authors discuss the stability of unparticles interacting with radiation.} Positive $\rho_u$ and $T_c$ require $-3\leq \delta \leq 0$ and $B<0$. This particular range of $\delta$ ensures that the evolution fulfills the Null Energy Condition. At early times unparticles are in thermal equilibrium with the SM, implying $T_{u} \sim T_{r}$, where $T_u$ and $T_r$ are unparticles and radiation temperature, respectively. At late times unparticles decouple and asymptote to a CC. This UDE scenario naturally resolves the fine-tuning problem of many DE models. The scenario also exhibits a built-in tracker mechanism \cite{Steinhardt:1999nw}, in which the energy density of unparticles tracks the radiation throughout the evolution, until it shifts relatively rapidly to a CC-like behavior, making the model immune to initial conditions problem. It is straightforward to write down the temperature ($y$), energy density ($\rho_u$) and equation of state ($w_u$) for unparticles at late-times , $y\gtrsim 1$  as a function of redshift: 
\begin{eqnarray}
y(z) &\simeq& 1 + \left( 1+ z\right)^{3}\,\left( y_0 -1 \right)\, , \label{eq:approxy(N)} \\
 \rho_{u}(z) &\simeq& -\frac{\delta  \sigma  T_c^4}{3 (\delta +4)} \left( 1 + \left( 1+ z\right)^{3}\, 4 (\delta +4)  \left(y_0-1\right)\right) \, , \label{eq:approxyrho(N)} \\
   w_{u} & \simeq & -1 + \left( 1+ z\right)^{3}\,\, 4 (\delta +4) \left(y_0-1\right) \, . \label{eq:approxwu}
\end{eqnarray}
\begin{figure}[]
\centering
\includegraphics[height=5.0cm]{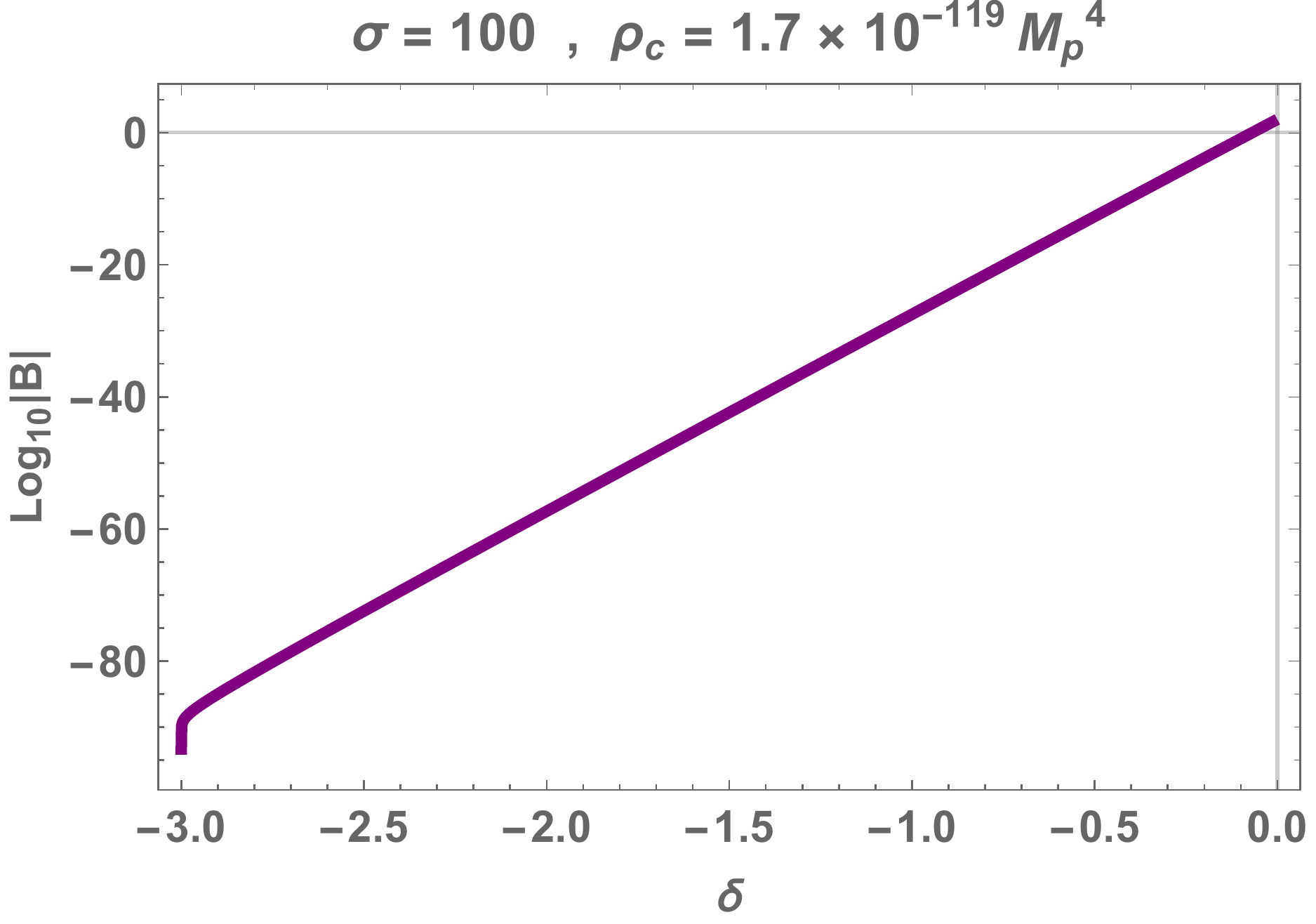} 
\includegraphics[height=5.0cm]{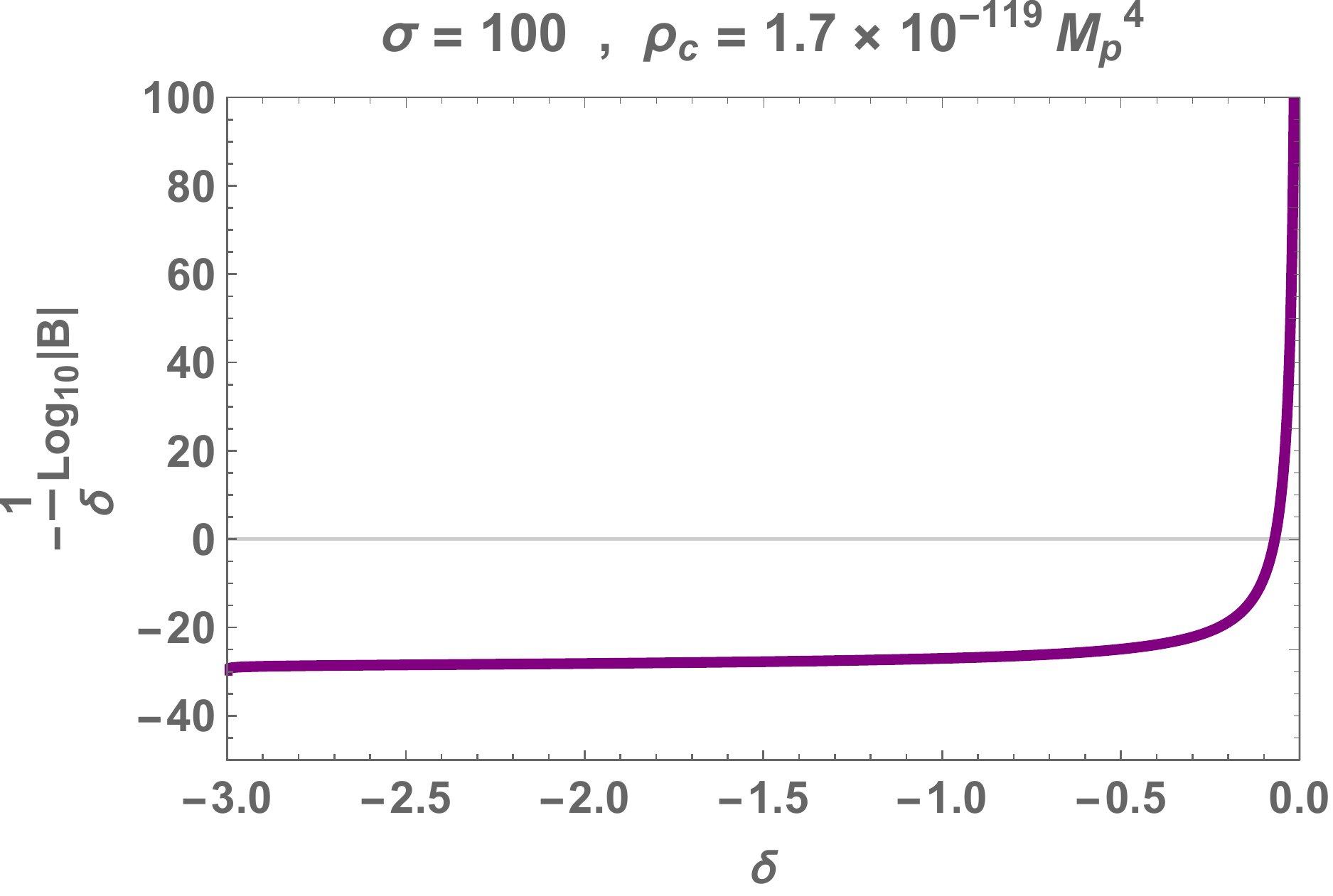}
\caption{\it 
 Left panel: The dependence of $B$ on $ \delta $ such that the energy density of unparticles accounts for the energy density of DE today $\Omega_{DE,0}$. Right panel: A comparison of B with respect to the Planck scale. B has dimensions of $M_{p}^{-\delta}$. Note that $\delta  \sim -0.068$ one obtains B is equal to the Planck scale. In all cases, we take $\sigma=100$.  } 
\label{fig:Bdelta}
\end{figure}
\begin{figure}[]
\centering
\includegraphics[scale=0.53]{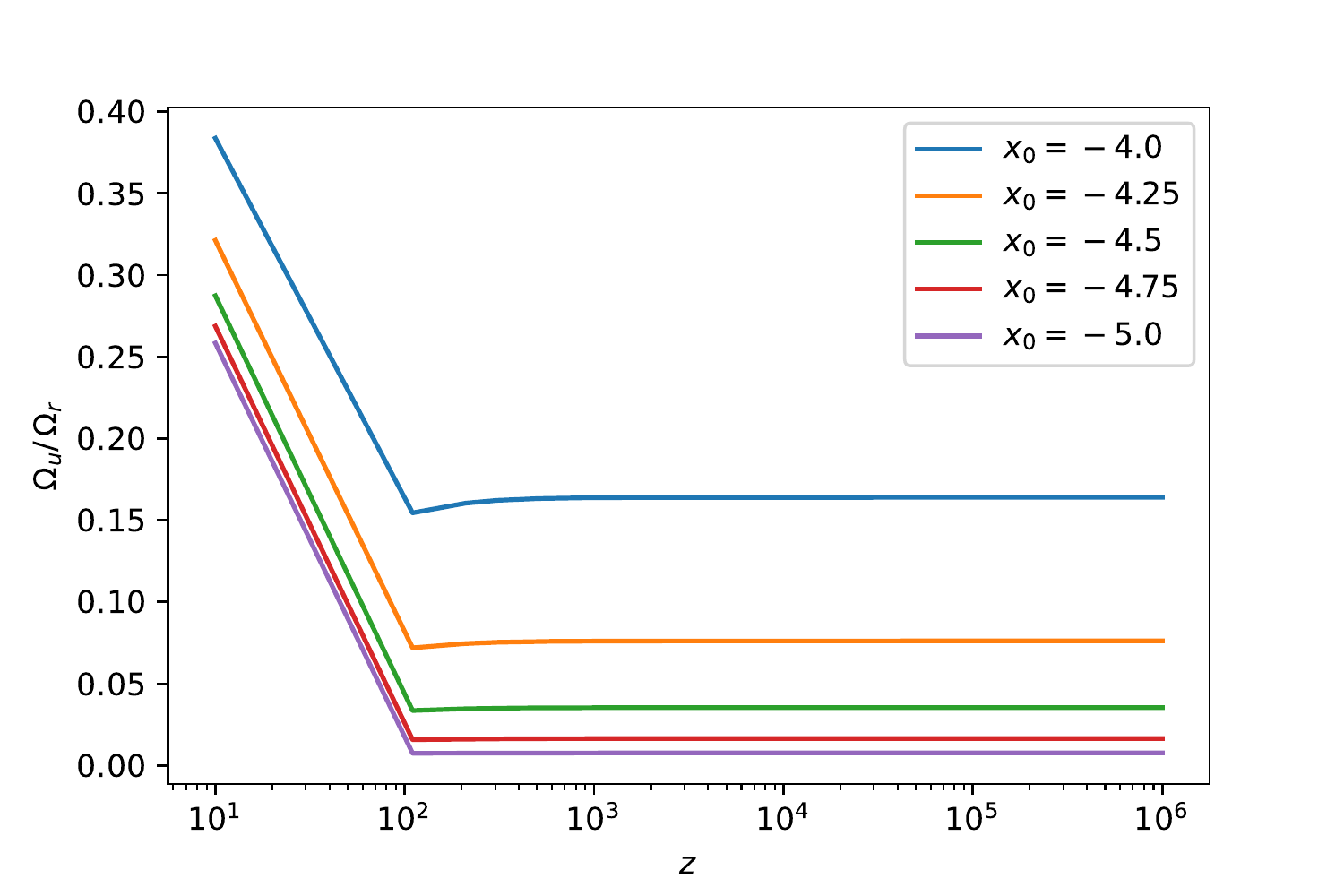} 
\includegraphics[scale=0.53]{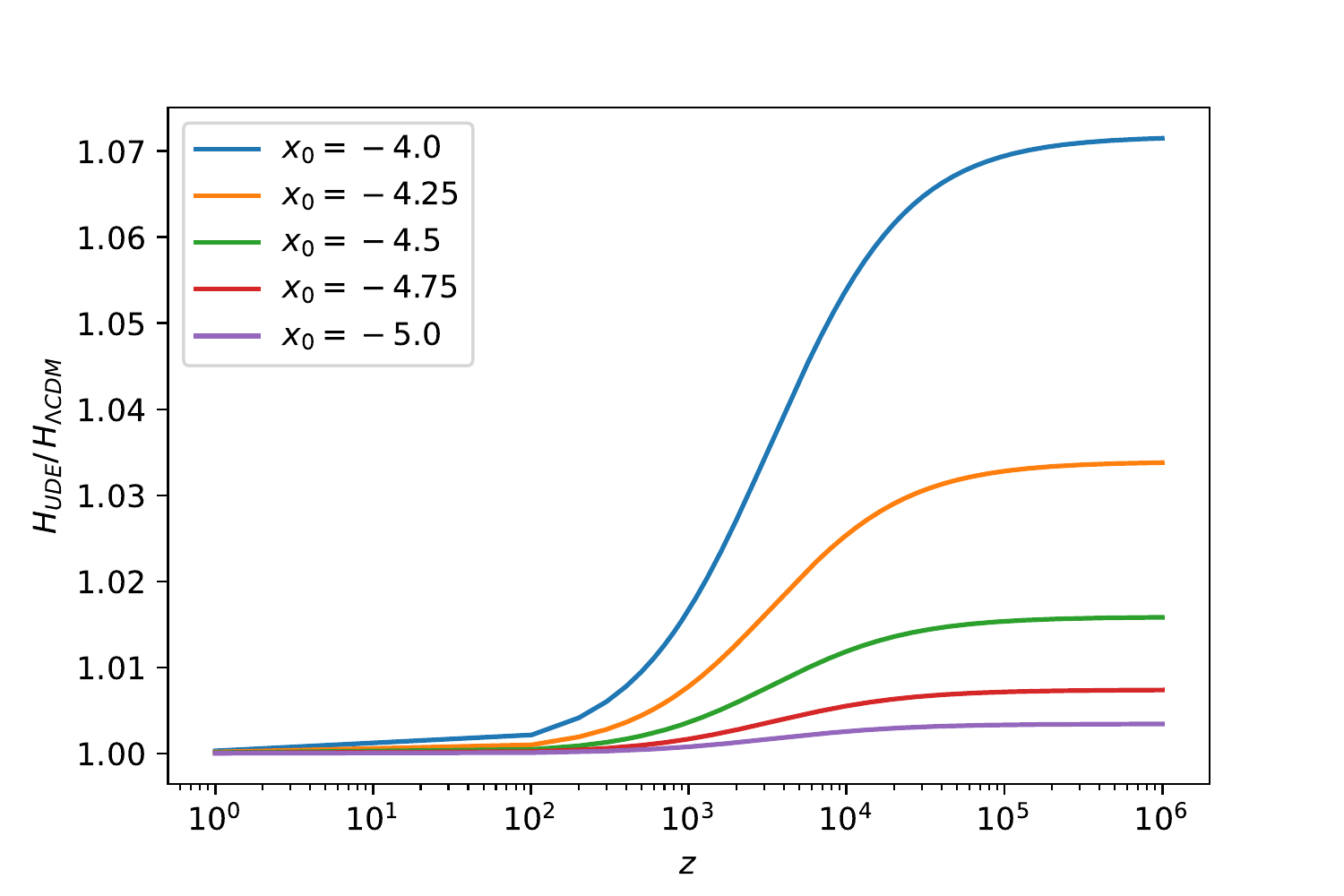}
\caption{\it 
 Left panel: Ratio of unparticles energy density to the radiation energy density for different values of $x_0 \in [-5.0, -4.0]$, where $y_0=1+10^{x_0}$. Right panel: The ratio between the Hubble parameter in UDE to that of $\Lambda$CDM as a function of redshift. Both plots show the different background behavior of the UDE model as a function of redshift. We have used the best-fit values for the cosmological parameters for Planck 2018 for both models given in Table \ref{tab:Planck 2018} in these plots.
} 
\label{fig:back}
\end{figure}

The connection between the energy density of DE according to the data \cite{Planck:2018vyg}, $\rho_c  \simeq 1.7 \times 10^{-119}\, M_p^4 $, and the parameters of the model is given by:
\begin{equation}
\rho_c = - \frac{\delta}{3\left( 4 + \delta\right)} \left( -\frac{4\, \sigma}{3\, B} \,\frac{\left(\delta + 3\right)}{\left( \delta + 4\right)}\right)^{\frac{4}{\delta}}
\end{equation}
 
In Fig. \ref{fig:Bdelta}, we illustrate the parameter space for permissible values of $B$ as a function of $\delta$, which reproduces the current energy density of DE. Considering $\sigma = 100$, notice that this energy scale $B^{-1/ \delta}$ could be in a huge span of energies, $10^{-30}\, M_p < B^{-1/ \delta} < M_p$. Finally, in Fig. \ref{fig:back}, we illustrate the evolution of the ratio of the energy density of unparticles to the energy density of the radiation content of the Universe (left panel) and the evolution of the ratio of the Hubble parameter UDE and $\Lambda$CDM cosmologies given the same initial value at $z=0$ in the right panel. The initial value has been taken as the $\Lambda$CDM one from Table \ref{tab:Planck 2018}, $H_0=67.3$ and the increase in the Hubble parameter as a function of redshift by a few percent in UDE is obvious.

\subsection{Perturbations}
Next, we move to the evolution of perturbations in the presence of unparticles. Perturbations to the FLRW background give rise to CMB and the structure that we see today. We study the evolution of linear perturbations using the publicly available Einstein-Boltzmann equation solver code \texttt{CAMB} \cite{Lewis:1999bs}. We modify the \texttt{CAMB} dark energy module in order to study the evolution of background and perturbation. We use the perfect fluid prescription of unparticles. In a spatially flat universe, the evolution of the density contrast $\delta_u$ and the fluid velocity $\theta_u$ of unparticles are governed by the following equations \cite{Ma:1995ey}.
\begin{eqnarray}
    \delta_{u}' &=& -\left( 1 + w_u \right) (\theta_{u} - 3\,\Phi') - \frac{a'}{a} \, \left( \frac{\delta p_u}{\delta \rho_u} - w_{u}\right)\,\delta_{u} \,,\nonumber \\
    \theta_{u}' &= & -\frac{a'}{a}\, \left( 1 - 3\,w_{u}\right)\,\theta_u - \frac{w_{u}'}{1 + w_{u}} \,\theta_u + \frac{\frac{\delta p_u}{\delta \rho_u}}{1 + w_u} \,k^2\,\delta_{u} + k^2\,\Phi  \label{eq:perturbations}
\end{eqnarray}
where $\frac{\delta p_u}{\delta \rho_u}$ is the adiabatic sound speed and $\Phi$ is the gravitational potential. These equations are solved numerically while assuming adiabatic initial conditions.

\begin{figure}[]
\centering
\includegraphics[scale=0.53]{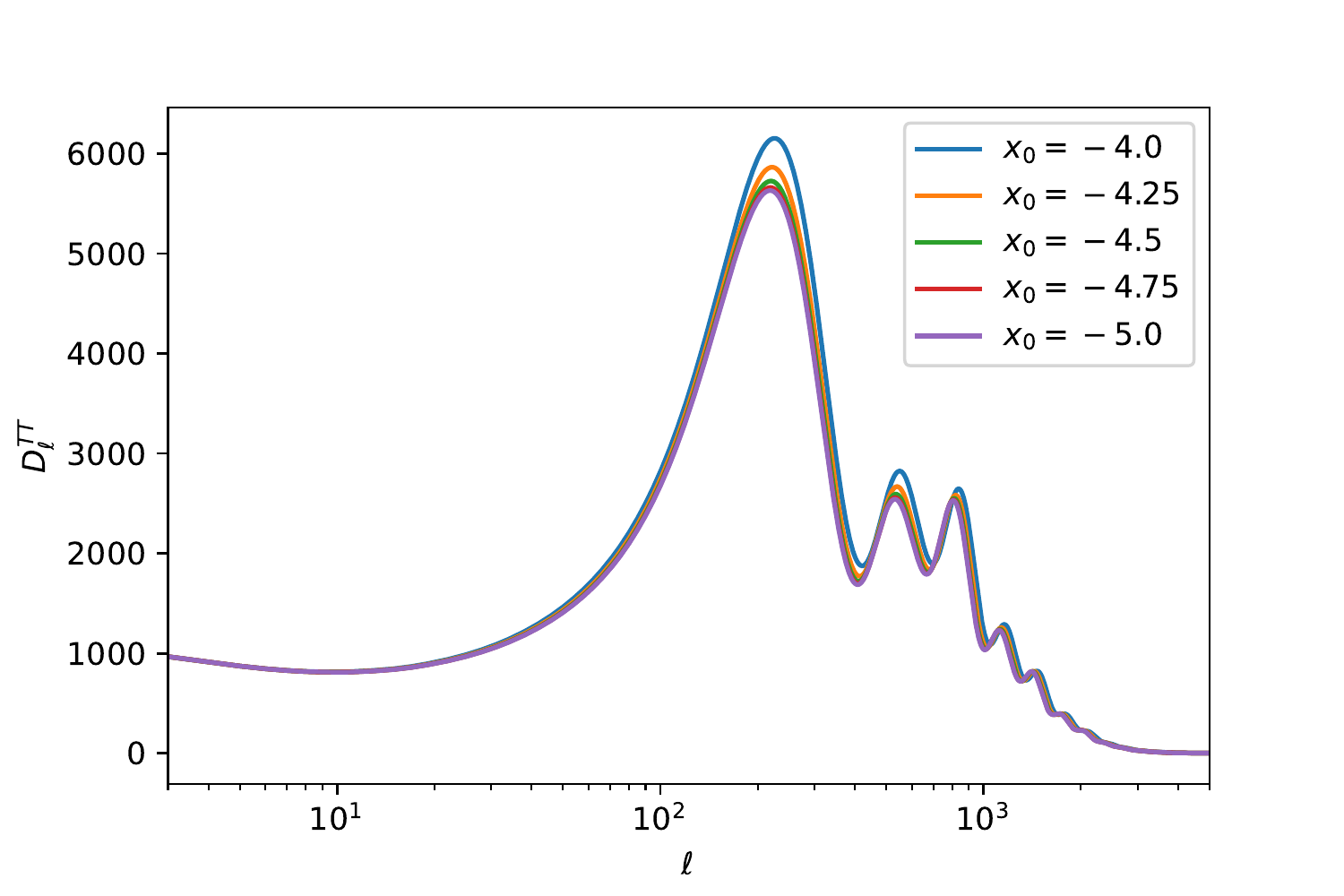} 
\includegraphics[scale=0.53]{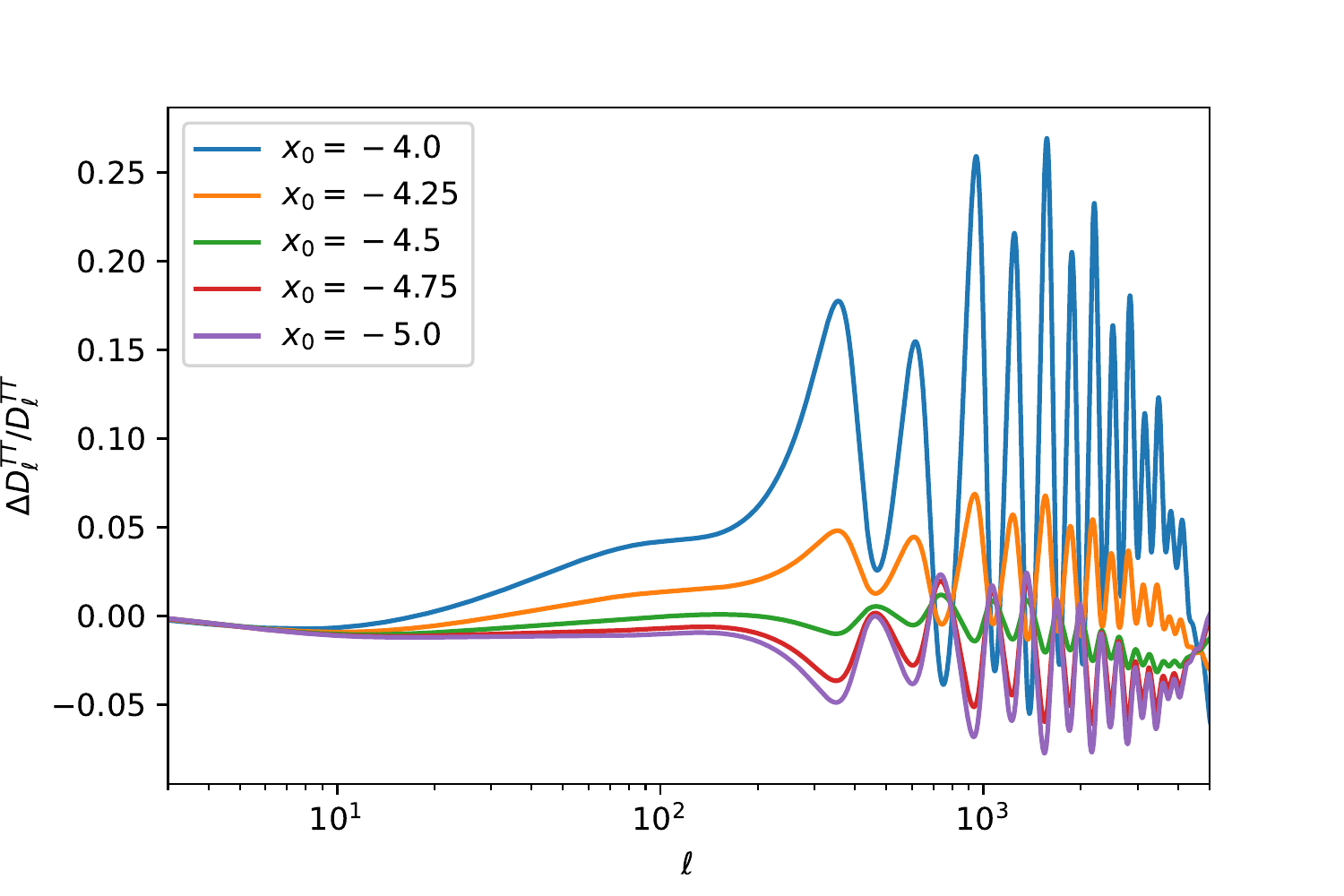}
\caption{ Left Panel: Plot of CMB temperature anisotropy power spectra for different $x_0$. The peaks of the temperature spectrum $D_{\ell}^{TT}$ are suppressed as we decrease $x_0$. Right Panel: Residuals $\Delta D_{\ell}^{TT}$ for UDE with respect to $\Lambda$CDM using the best fit values of $H_0 = 69.99\, \text{km/sec/Mpc}$ and $H_0 = 67.3\, \text{km/sec/Mpc}$, respectively.
 } 
\label{fig:cll}
\end{figure}

Hence, we can analyze the effect of UDE on CMB and matter power spectra. In Fig (\ref{fig:cll}) we show the CMB temperature spectrum $D_{\ell}^{TT}=[\ell (\ell+1)]^2 C_{\ell}^{TT}/2\pi$ for different values of $x_0$ and the fractional change $\Delta D_{\ell}^{TT}$. We choose the best-fit values of cosmological parameters from the \texttt{Planck 2018} derived for both models. It is clear from the left panel in Fig \ref{fig:cll} that a decrease in the present temperature of unparticles leads to suppression in peaks of $D_{l}^{TT}$ while the right panel shows the amplification in $D_{l}^{TT}$ for $\ell \in (10 - 1000)$ and oscillations in residual of $D_{l}^{TT}$ for large $\ell > 10^{3}$. It will be interesting to understand these amplifications and oscillations as late or early-time integrated Sachs-Wolfe (ISW) signals  \cite{Krolewski:2021znk,Vagnozzi:2021gjh,Cabass:2015xfa}.

\begin{figure}[]
\centering
\includegraphics[scale=0.53]{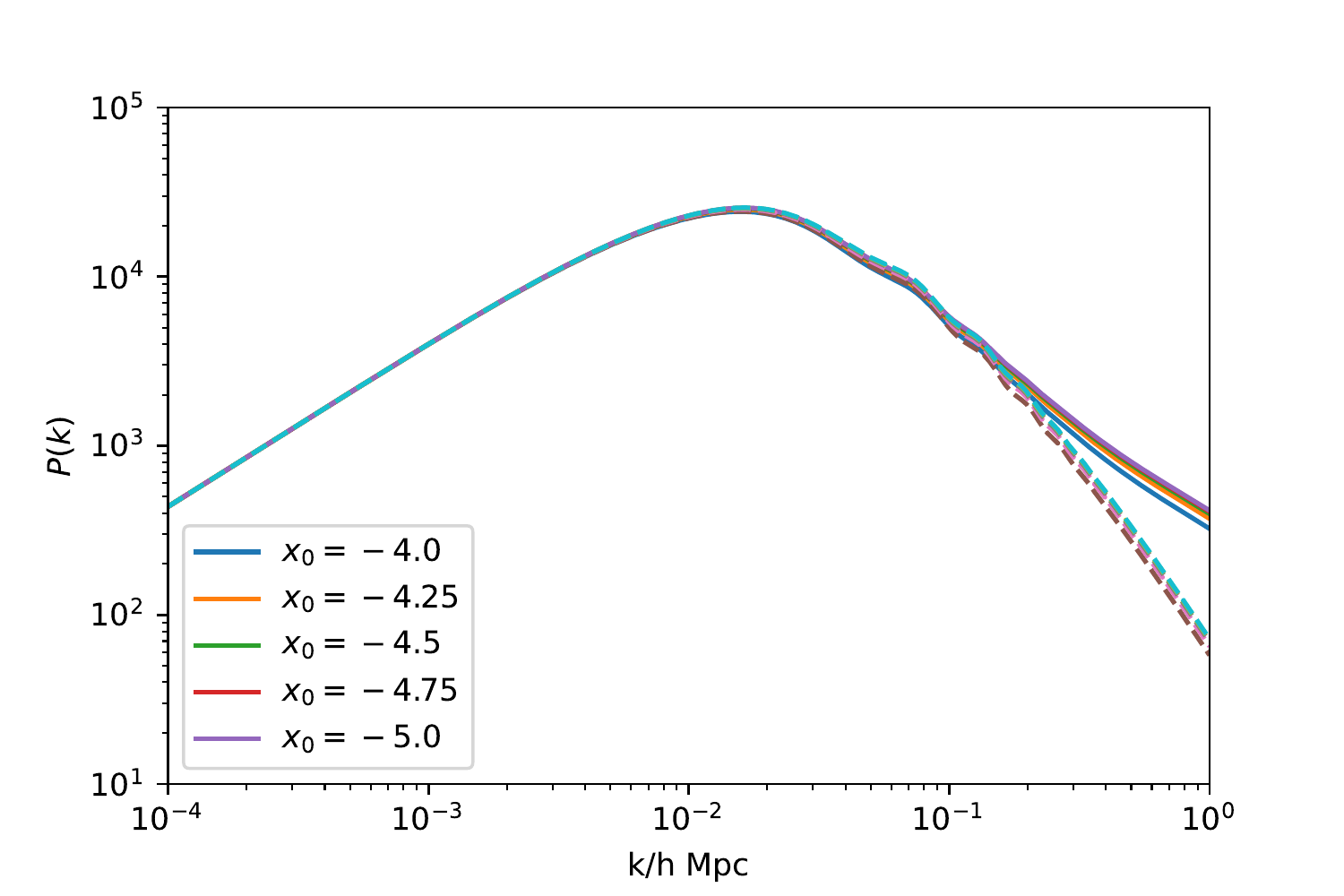} 
\includegraphics[scale=0.53]{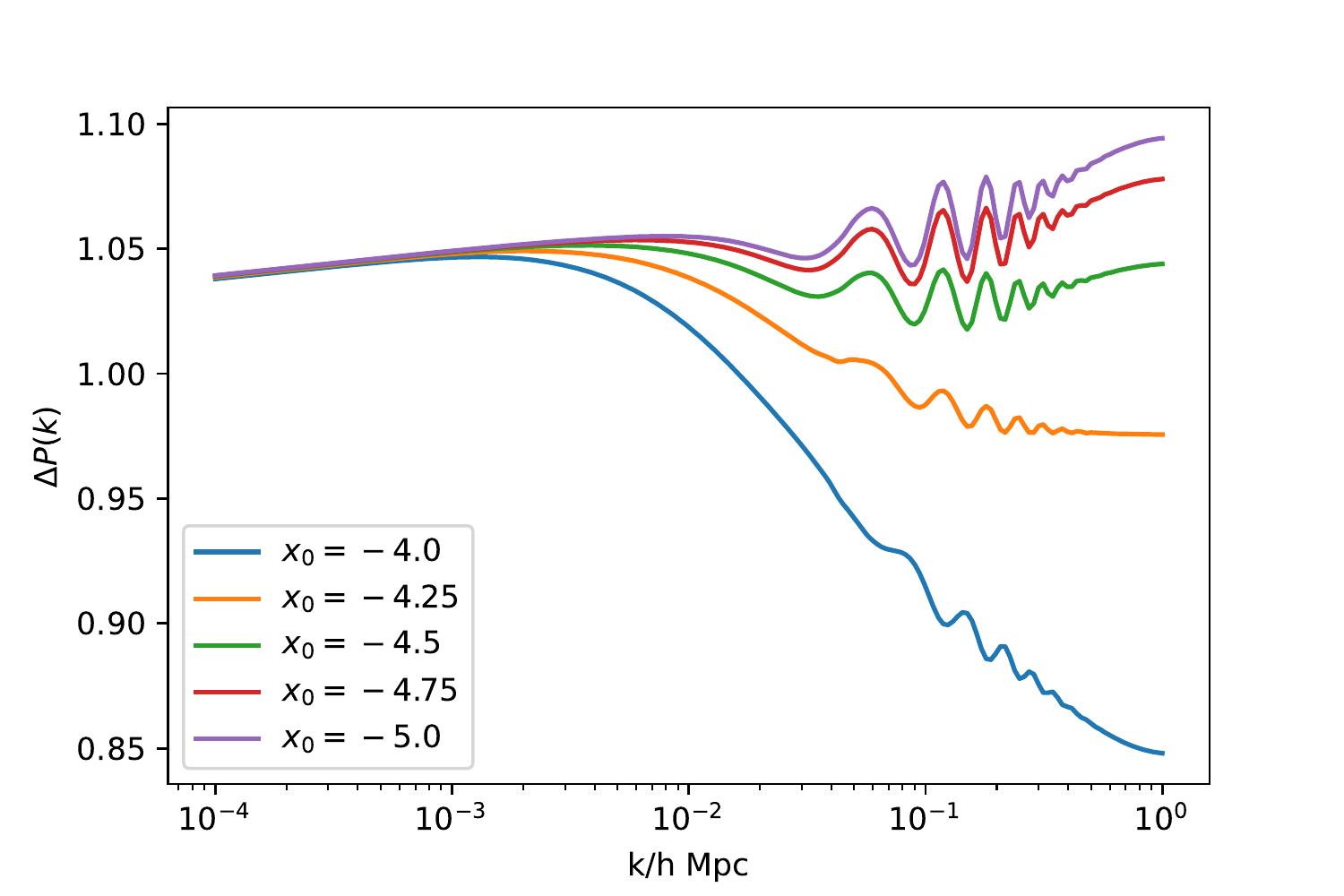}
\caption{ Left Panel: The Non-linear (solid) and linear (dashed) matter power spectrum $P(k)$ at $z = 0$ for the UDE model that fits the primary CMB data. The amplitude of $P (k)$ in the range $0.1 \,h / \text{Mpc} \leq k \leq 1 h /\,\text{Mpc}$ decreases for higher temperature of unparticles today. Right panel: Ratio of UDE and $\Lambda$CDM non-linear matter power spectra at $z = 0$. The model parameters are the same as in the previous figures. The suppression in the amplitude of $P(k)$ has a significant role in the change of $\sigma_8$ and hence $S_8$. This decrease in $S_8$ is the result of shifts in standard cosmological parameters in the UDE model.} 
\label{fig:pk}
\end{figure}

We then investigate the UDE affecting observables beyond the CMB temperature and polarization power spectra. The left panel of Fig \ref{fig:pk} shows the increase in the nonlinear matter power spectrum as a result of a decrease in unparticles' temperature. The right panel again shows the departure of the matter power spectrum of the UDE scenario with respect to $\Lambda$CDM. We see the oscillatory departure from the $\Lambda$CDM at small scales. 

\section{Data} \label{sec:data}
The inferred values of the cosmological parameters have certain dependence on the data sets used. Different combinations of data sets will result in somewhat different values. We use various data sets to constrain the UDE model and compare the results to $\Lambda$CDM. We modify the publicly available Einstein-Boltzmann code \texttt{CAMB} \cite{Lewis:1999bs} in conjunction with \texttt{Cobaya} \cite{Torrado:2020dgo} to perform the Markov-Chain Monte Carlo (MCMC) simulations. In our analysis, we use the following publicly available data sets: 

\begin{itemize}
\item{ Planck 2018 CMB :} First, we consider the \texttt{Planck 2018} likelihood for the CMB data, which consists of the low-l TT, low-l EE, and high-l TTEETE power spectra \cite{Planck:2019nip}. We also use the \texttt{Planck 2018} lensing likelihood \cite{Planck:2018lbu}, which has an important role in the LSS analysis of the Late Universe. 

\item{Baryon Acoustic Oscillations (BAO) and RSD measurements:} We use the measurements from the SDSS DR7 Main Galaxy Sample (MGS) \cite{Ross:2014qpa} and 6dF galaxy survey \cite{Beutler:2011hx} measurements at $z = 0.15$ and $z = 0.106$ respectively. In addition to that, we also include BAO and $f\,\sigma_8$ measurements (where f is the linear growth rate) from BOSS DR12 $\&$ 16 at $z = 0.38,0.51,0.68$ \cite{BOSS:2016wmc,Bautista:2020ahg,Gil-Marin:2020bct,eBOSS:2020yzd}, QSO measurements at $z = 1.48$ \cite{Neveux:2020voa,Hou:2020rse} and Ly-$\alpha$ auto-correlation and cross-correlation with QSO at $z = 2.2334$ \cite{duMasdesBourboux:2020pck}. 

\item{DES:} Dark Energy Survey includes measurements from shear-shear, galaxy-galaxy, and galaxy-shear two-point correlation functions, referred to as "$3 \times 2$ pt", measured from 26 million source galaxies in four redshifts bins and 650,000 luminous red lens galaxies in five redshifts bins, for the shear and galaxy correlation functions \cite{DES:2021wwk}. DES $3\times2$ pt likelihood gives $S_8 = 0.773^{+0.026}_{-0.020}$ and $\Omega_m = 0.267^{+0.030}_{-0.017}$  for $\Lambda$CDM model.

\item{Supernovae Pantheon:} The Pantheon data set is a collection of the absolute magnitude of 1048 supernovae distributed in redshift interval $0.01 < z < 2.26 $ \cite{Pan-STARRS1:2017jku}. Many times we will simply refer to this data set as SN.

\item{$H_0$ from \texttt{SH0ES}:} We use latest local measurement of $H_0 = 73.04 \pm 1.4 \text{km/sec/Mpc}$ from the \texttt{SH0ES} team \cite{Riess:2020fzl}. Many times we will simply refer to this data set as $H_0$.
\end{itemize}

\section{Constraints on Unparticles Dark Energy}  \label{sec:results}
\subsection{Priors}
Let us compare the cosmological parameter constraints on UDE and $\Lambda$CDM. We fit the different data sets described in the previous section alone or several of them combined together. We fix the parameter $\delta=-3$ because it allows quick implementation of the model in CAMB. However, we have verified that any value of $\delta $ lying in the range $[-3,0]$ will not considerably affect the cosmological observables using compressed likelihoods \cite{Zhai:2019nad}. The other free parameter of the UDE model is the present value of unparticles dimensionless temperature $y_0$, which we set free. For numerical implementation, we parameterize $y_0$ as $1 + 10^{x_0}$ where $y_0 = 1$ shows the temperature of unparticles at future infinity and adopt a uniform prior $x_0 \in [-4.5,-3]$. Except for $x_0$ we constrain the other standard cosmological parameters for both cosmologies - the baryon matter density $\Omega_b h^{2}$, the cold dark matter density $\Omega_{c}h^{2}$, the amplitude of primordial curvature spectrum amplitude $A_s$ evaluated at suitable pivot scale, $k = 0.05 Mpc^{-1}$ along with its tilt $ n_s $, and the reionization optical depth $\tau_{reio}$. We use the standard three neutrino description with one massive with mass, $m_{\nu}$ = 0.06 eV, and two massless neutrinos. Table \ref{tab:param_prior} lists the priors for different parameters for all cosmologies described above. In each subsection, we explain the reasons for choosing the particular combination of data sets.

\begin{table}
\caption{Priors used on various free parameters of $\Lambda$CDM and Unparticles model* for MCMC analysis}
   \vspace{1 em} 
\centering
\begin{tabular}{|c|c|}

\hline
\hline

    Parameter         & Prior  \\
\hline
\hline
$\Omega_b h^2$          & [0.005, 0.1]    \\
$\Omega_c h^2$          & [0.001,0.99]  \\
$H_0$        & [20,100]  \\
$\tau_{reio}$                  & [0.01,0.8]  \\
${\rm{ln}}(10^{10} A_s)$& [1.6,3.9]  \\
$n_s$                   & [0.8,1.2]   \\
$ x_{0}$*               & [-4.5,-3]  \\
\hline
\hline
\end{tabular}
 \label{tab:param_prior}
\end{table}
\subsection{Results}

\subsubsection{\texttt{Primary \textit{Planck} 2018} } 
\begin{figure}[]
\centering
\includegraphics[scale=0.35]{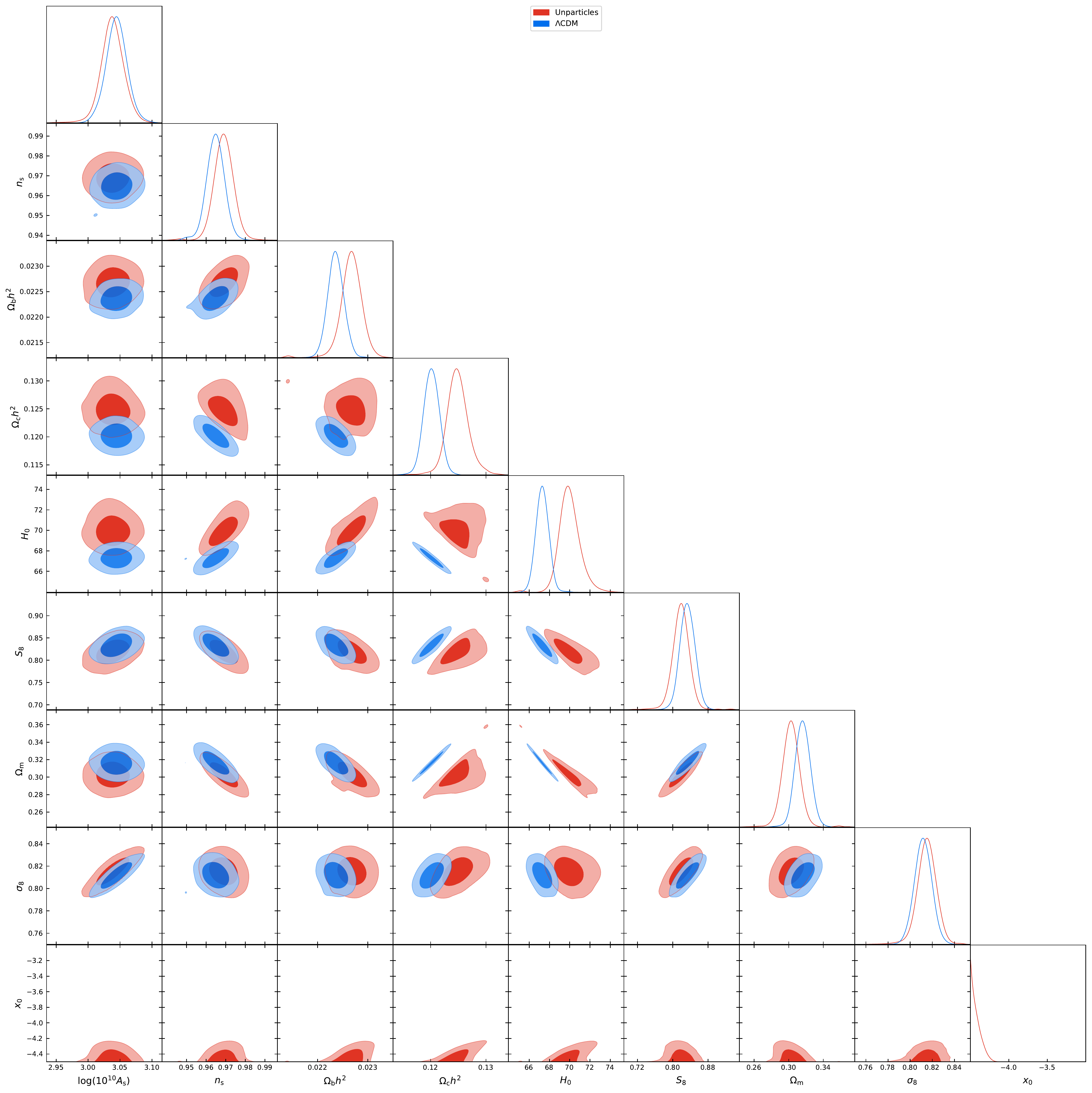}
\caption{ Posterior distribution of cosmological parameters \textit{Planck} 2018 CMB data only (\texttt{TTEETE}) for $\Lambda$CDM and UDE scenario.  }
\label{fig:planck2018}
\end{figure}

We first consider the \textit{Planck} 2018 primary CMB TT, TE, and EE power spectrum data, as it is one of the main drivers of the tension. We have analyzed the Pantheon data only in \cite{Artymowski:2021fkw}. The constraints on cosmological parameters at 1-$\sigma$ CL are tabulated in Table \ref{tab:Planck 2018}. We find a bound on $x_0 < -4.40$ at 68 \% CL. A comparison of the posterior distributions in UDE and $\Lambda$CDM is presented in Fig \ref{fig:planck2018}. We find the Hubble constant in UDE to be $H_0 = 69.99^{+0.84}_{-1.1}$ km/sec/Mpc, shifted upwards compared to $\Lambda$CDM for the same data, $H_0 = 67.30\pm 0.65$ km/sec/Mpc with slightly larger error bars. We also find $S_8 = 0.819\pm 0.021$ with significantly lower value than the $\Lambda$CDM value $S_8 = 0.833\pm0.017$. However, this value of $S_8$ is still larger than the DES-only result. The possible reason for reducing the Planck derived $S_8$ value towards the lower $S_8$ from DES is the reduction of total matter density ($\Omega_m$) to $0.303\pm 0.011$ in UDE from the $\Lambda$CDM scenario $\Omega_m = 0.3162\pm 0.008$ while keeping the $\sigma_8$ value almost same. 

\begin{table}[H]  
\textbf{Constraints from \textit{Planck} 2018 CMB}
   \vspace{1 em}  \\
\centering
\begin{tabular}{|c|c|c|c|}

\hline
\hline

    Parameter         & $\Lambda$CDM &  UDE  \\
\hline
\hline
$\ln(10^{10} A_\mathrm{s})$        & $3.045\pm 0.017 \,(3.042)$&  $3.037\pm 0.019\,(3.038)$  \\
$n_\mathrm{s}$          & $0.9647\pm 0.0047 \, (0.9649)$& $0.9689\pm 0.0054\,(0.968)$ \\
$\Omega_b h^2$        & $0.02236\pm 0.00016\, (0.02234)$  & $0.02267\pm 0.00023\,(0.0226)$ \\
$\Omega_c h^2$          & $0.1202\pm 0.0014 \, (0.1204)$& $0.1249^{+0.0017}_{-0.0019}\,(0.1240)$ \\
$\tau_{reio}$                  & $0.0541\pm 0.0082 \, (0.05268)$ & $0.0559\pm 0.0091\,(0.0538)$ \\
 Age & $13.799\pm 0.027 \,(13.80)$ & $13.43^{+0.10}_{-0.049}\,(13.52)$\\
$ x_0$          & -& $< -4.40\,(-4.49)$ \\
\hline
\hline
$ H_0 $               & $67.30\pm 0.65 \,(67.14)$& $69.99^{+0.84}_{-1.1}\,(69.14)$ \\
$\sigma_8$               & $0.8117\pm 0.0079\,(0.8128)$& $0.815\pm 0.010\,(0.814)$ \\
$ S_8$               & $0.833\pm 0.017\,(0.836) $&  $0.819\pm 0.021\,(0.825)$\\
$ \Omega_m$               & $0.3162\pm 0.0089 \,(0.3182)$& $0.303\pm 0.011\,(0.308)$ \\
\hline 
    \hline
         low-$\ell$ TT & 23.32 &22.67\\
         low-$\ell$ EE &395.85 & 395.98\\
     high-$\ell$ TTTEEE &  2344.99    & 2347.65\\

    \hline
    Total $\chi^2 $   & 2764.17 & 2766.31\\
     $\Delta \chi^2 $ & 0  & 2.14\\ 

\hline
\hline

\end{tabular}
 \caption{ The mean $\pm 1 \sigma$(best-fit) constraints on the cosmological parameters inferred from the \textit{Planck} 2018 CMB data only (\texttt{TTEETE}) for $\Lambda$CDM and UDE scenario. The constraints are reported at the 68 \% CL. We also report the $\chi^{2}_{min}$ for each model and data sets. We find that $H_0 $ and $S_8 $ are reduced to $1.7 \sigma$ and $1.37 \sigma$ with $\texttt{SH0ES}$ and \texttt{DES} measurements respectively. We find an upper bound on unparticles temperature today $x_0 < -4.40$.  } \label{tab:Planck 2018}
\end{table}
The goodness-of-fit to the primary CMB anisotropies, quantified by the $\chi^{2}$-statistic, is worsened in the UDE scenario with one extra parameter, yielding $\Delta\,\chi^{2} = \chi^{2}_{UDE} - \chi^{2}_{\Lambda CDM}=2.14$. 

\subsubsection{\texttt{Primary \textit{Planck} 2018}+\texttt{Lensing}+\texttt{BAO}+\texttt{SN}+\texttt{$H_0$}}
\begin{figure}[]
\centering
\includegraphics[scale=0.35]{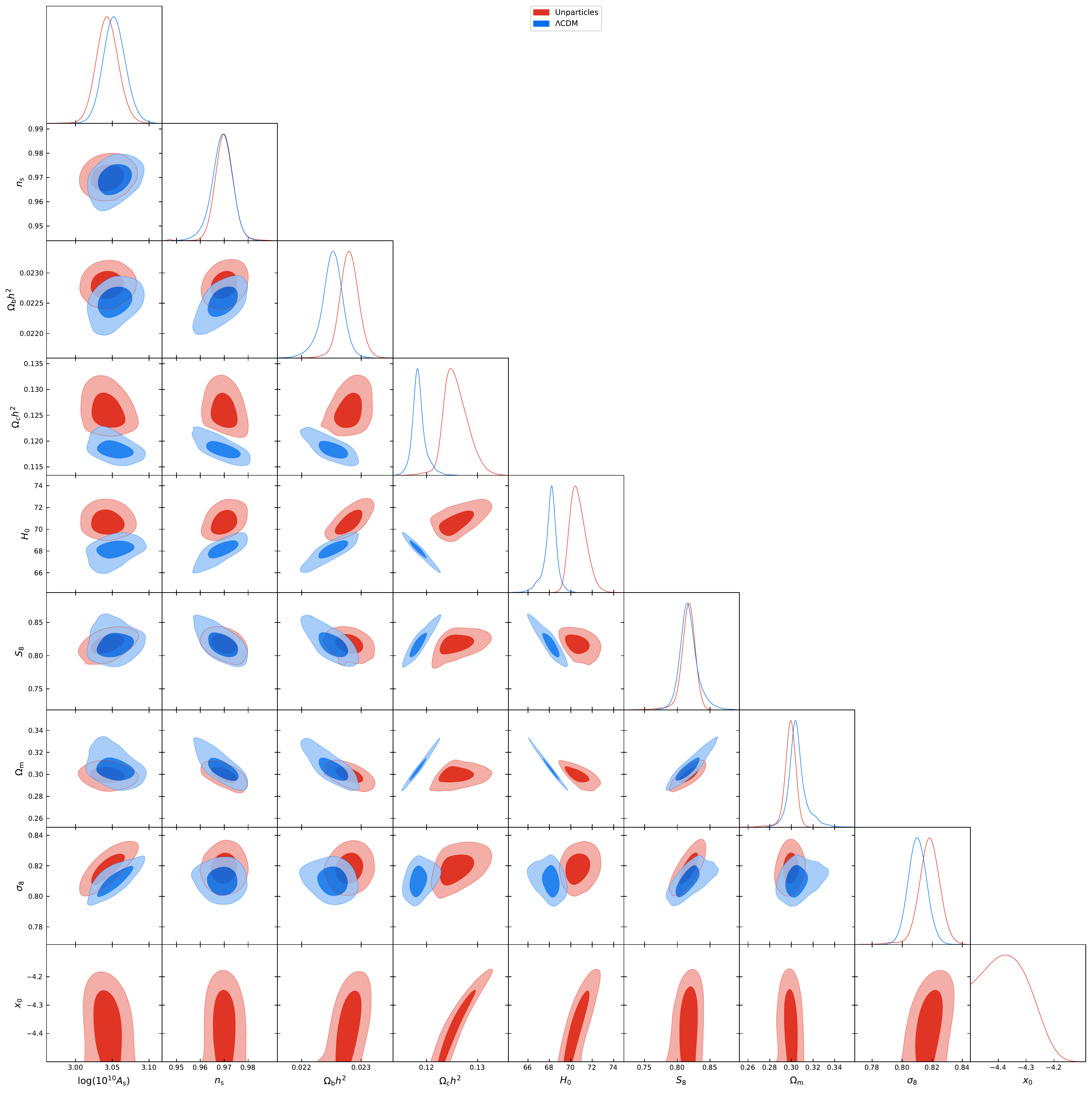}
\caption{Posterior distribution of cosmological parameters using \textit{Planck} 2018 CMB (\texttt{TTEETE}) +\texttt{Lensing}+\texttt{BAO}+\texttt{SN}+\texttt{SHOES} for $\Lambda$CDM and UDE scenario. We present the contour plots for $\Lambda$CDM and UDE in blue and red respectively with $1 \sigma$ and $2 \sigma$ CL.}
\label{fig:Planck+Lensing+BAO+SN+H0}
\end{figure}

We now combine the Primary \textit{Planck} 2018 with \textit{Planck} 2018 CMB lensing, BAO, Pantheon, and SH0ES likelihood, such that it can be compared to the DES result that its $S_8$ measurement is in tension with Planck. The Posterior distributions for this combination of data sets are shown in Fig \ref{fig:Planck+Lensing+BAO+SN+H0}. The best-fit parameters and $ 68 \%$ CL constraints on cosmological parameters are tabulated in Table \ref{tab:Planck+Lensing+BAO+SN+H0}.  We find $H_0 = 70.69^{+0.60}_{-0.89}$ km/sec/Mpc reducing the $H_0$ tension with \textsc{SH0ES} to $\approx 1.39\,\sigma$ from the $H_0 = 68.09^{+0.63}_{-0.41}$ inferred from $\Lambda$CDM which is in $3.19 \sigma$ tension with the SH0ES result.
\begin{table}[H]
\textbf{ Constraints from \textit{Planck} 2018 CMB+Lensing+BAO+SN+$H_0$}
\vspace{2 em} \\
\centering
\begin{tabular}{|c|c|c|c|}

\hline
\hline

    Parameter         & $\Lambda$CDM & UDE \\
    
\hline
\hline
$\ln(10^{10} A_\mathrm{s})$          & $3.053\pm 0.015\,(3.0506)$ &  $3.043\pm 0.016 \,(3.045)$  \\
$n_\mathrm{s}$          & $0.9690^{+0.0046}_{-0.0038} \,(0.9704)$& $0.9700\pm 0.0043 \,(0.9707)$ \\

$\Omega_b h^2$        & $0.02251^{+0.00019}_{-0.00015}\,(0.02256)$ & $0.02280\pm 0.00016 \,(0.0228)$\\
$\Omega_c h^2$          & $0.11847^{+0.00088}_{-0.0013}\,(0.11881)$& $0.1258^{+0.0018}_{-0.0027}\,(0.1256)$ \\
$\tau_{reio}$                  & $0.0598\pm 0.0081\,(0.0591)$& $0.0592\pm 0.0079\,(0.06014)$ \\
 Age & $13.771^{+0.022}_{-0.029}\,(13.76)$ & $13.35^{+0.14}_{-0.073}\,(13.36)$\\
$ x_0$               & -& $-4.366^{+0.065}_{-0.10}\,(-4.366)$ \\
\hline
\hline
$ H_0 $               & $68.09^{+0.63}_{-0.41}\,(68.26)$& $70.69^{+0.60}_{-0.89}\,(70.55)$ \\
$\sigma_8$               & $0.8102\pm 0.0063\,(0.8128)$& $0.8179\pm 0.01\,(0.819)$ \\
$ S_8$               & $0.818^{+0.010}_{-0.015}\,(0.8128)$& $0.816^{+0.011}_{-0.0083}\,(0.818)$ \\

$ \Omega_m$               & $0.3056^{+0.0051}_{-0.0083}\,(0.3033)$& $0.2988^{+0.0055}_{-0.0041}\,(0.30)$ \\
\hline 
\hline
\textit{Planck} 2018 & 2767.08 &2768.13\\
\textit{Planck} Lensing &8.84 & 9.13\\
 BAO &29.00 &  28.93\\
 SN&  1034.77  &   1034.73\\
 $H_0$ &14.40 &  4.16\\

    \hline
    Total $\chi^2 $   & 3854.11 & 3845.10\\
     $\Delta \chi^2 $ & 0  & -9.01\\ 
\hline
\hline
\end{tabular}

\caption{ The mean $\pm 1 \sigma$(best-fit) constraints on the cosmological parameters inferred from the \textit{Planck} 2018 CMB data (\texttt{TTEETE})+\texttt{Lensing}+\texttt{BAO}+\texttt{SN}+\texttt{$H_0$} for $\Lambda$CDM and the UDE scenario. We also report the $\chi^{2}_{min}$ for each model and data sets. We find evidence of detecting UDE with $x_0 = -4.37^{+0.07}_{-0.10}$. The UDE scenario is favoured by this combination of data set with $\Delta \chi^{2} = -9.01$ with respect to $\Lambda$CDM.}  \label{tab:Planck+Lensing+BAO+SN+H0}
\end{table}
Considering the LSS tension, there is little difference as 
the value of $S_8$ is brought down with $S_8 = 0.816^{+0.011}_{-0.008}$ for UDE compared to $S_8 = 0.818^{+0.010}_{-0.015}$ for $\Lambda$CDM.
We find $x_0 =-4.366^{+0.065}_{-0.10}$, and $\Delta \chi^{2} = -9.01$, which clearly shows evidence of UDE using this particular combination of data. The improvement in $\Delta\chi^2$ is dominated by the improvement with respect to the SH0ES data $\Delta\,\chi^2_{\texttt{SH0ES}} = -10.24$.

\subsubsection{\texttt{Primary Planck 2018}+\texttt{Lensing}+\texttt{BAO}+\texttt{SN}+\texttt{DES}} 

\begin{table}[H]
\textbf{Constraints from \textit{Planck} 2018 CMB+Lensing+BAO+SN+DES}
   \vspace{1 em} \\
\centering
\begin{tabular}{|c|c|c|c|}

\hline
\hline

    Parameter         & $\Lambda$CDM & UDE \\ 
\hline
\hline
$\ln(10^{10} A_\mathrm{s})$          & $3.048\pm 0.015$\,(3.044) &    $3.039\pm 0.017$\,(3.0432)\\
$n_\mathrm{s}$          & $0.9691\pm 0.0036$\,(0.9695)&  $0.9696^{+0.0040}_{-0.0036}$\,(0.9709)\\
$\Omega_b h^2$        & $0.02253\pm 0.00014$ \,(0.02253) & $0.02269^{+0.00021}_{-0.00014}$\,(0.0227)\\
$\Omega_c h^2$          & $0.11813\pm 0.00080$\,(0.1182)&  $0.1238^{+0.0013}_{-0.0016}$\,(0.1227)\\
$\tau_{reio}$                  & $0.0580\pm 0.0078$\,(0.05614)&  $0.0569\pm 0.0091$\,(0.0586)\\
 Age & $13.768\pm 0.020$\,(13.76)  & $13.438^{+0.085}_{-0.040}$\,(13.48)\\
$ x_0$               & -& $<-4.42$ \,(-4.49) \\
\hline
\hline
$ H_0 $               & $68.22\pm 0.36$\,(68.186)&  $70.22^{+0.49}_{-0.66}$\,(69.92)\\
$ \sigma_8$               & $0.8072\pm 0.006$\,(0.8062)& $0.8123^{+0.0068}_{-0.0062}$ \,(0.8128)\\
$ S_8$               & $0.8122\pm 0.0093$\,(0.8117)&  $0.810^{+0.011}_{-0.0080}$\,(0.8114)\\
$ \Omega_m$               & $0.3037\pm 0.0047$\,(0.3041)& $0.2985^{+0.0054}_{-0.0044}$ \,(0.299)\\

\hline
\hline
\textit{Planck}  & 2775.47 &2776.10\\
BAO &29.14 & 28.30\\
DES &  509.11   & 509.15\\
SN &  1034.79    & 1034.73\\
\hline
Total $\chi^2 $   & 4348.52 & 4349.29\\
$\Delta \chi^2 $ & 0  & 0.77 \\
\hline
\hline
\end{tabular}
 
\caption{ The mean $\pm 1 \sigma$(best-fit) constraints on the cosmological parameters inferred from the \textit{Planck} 2018 CMB data (\texttt{TTEETE})+ \texttt{Lensing+BAO+SN}+\texttt{Dark Energy Survey -Y1} for $\Lambda$CDM and UDE.}
\label{tab:Base+Des}
\end{table}
\begin{figure}[]
\centering
\includegraphics[scale=0.35]{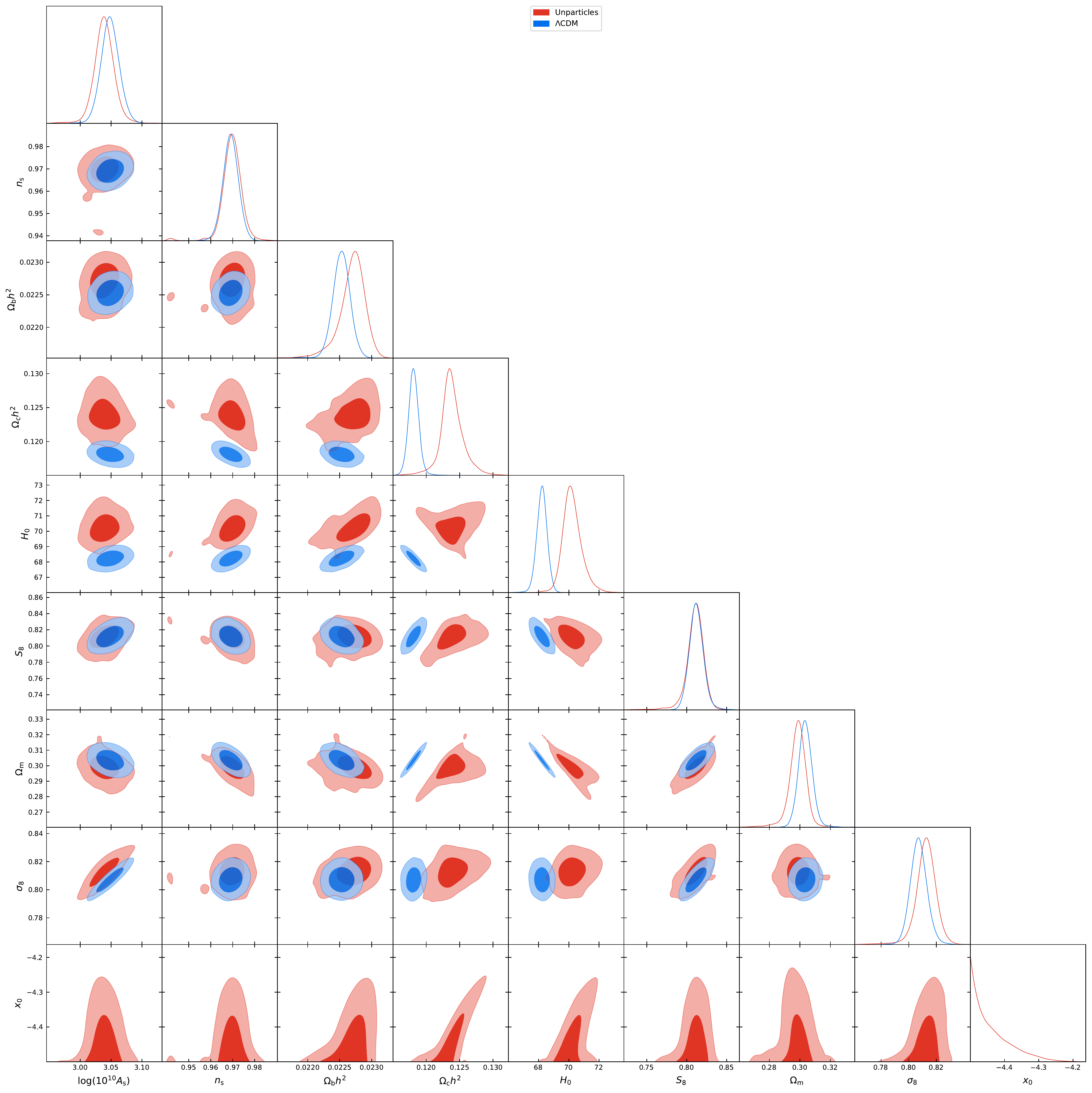}
\caption{Posterior distribution of cosmological parameters from the \textit{Planck} 2018 primary CMB+Lensing+BAO+SN+DES. We present the contour plots for $\Lambda$CDM and UDE in blue and red respectively with $1 \sigma$ and $2 \sigma$ CL.}
\label{fig:Planck+BAO+SN+DES}
\end{figure}

We now expand our analysis to include the \texttt{DES-Y1} data set. We jointly analyze \texttt{\textit{Planck} 2018 CMB}, \texttt{Lensing}, \texttt{BAO} and Pantheon data along with \texttt{DES-Y1} as to compare with the SH0ES data that is responsible for the $H_0$ tension with other probes. We implement the \texttt{Halofit} formula \cite{Mead:2020vgs} to account for the non-linear matter clustering, which is important to model galaxy-galaxy weak lensing correlation functions. The posterior distributions for our analysis, including the full \texttt{DES Y1} likelihood, are shown in Fig \ref{fig:Planck+BAO+SN+DES}. Parameter constraints with $\chi^{2}$ values are tabulated in Table \ref{tab:Base+Des}. We report that the inclusion of the \texttt{DES Y1} does not show significant evidence for UDE. The data sets put a bound on $x_0 < -4.42$. 

\begin{table}[H]
\textbf{ Constraints from Lensing+BAO+DES+SN+$H_0$}
   \vspace{2 em} \\
\centering
\begin{tabular}{|c|c|c|c|}

\hline
\hline

    Parameter         & $\Lambda$CDM & UDE \\ 
\hline
\hline
$\ln(10^{10} A_\mathrm{s})$          &$3.139^{+0.046}_{-0.11} \,(3.15)$  & $3.151^{+0.077}_{-0.097}\,(3.191)$   \\
$n_\mathrm{s}$          & $0.957\pm 0.051\,(0.95)$& $0.926\pm 0.047\,(.96)$ \\
$\Omega_b h^2$        & $0.0294^{+0.0027}_{-0.0020}\,(0.0297)$  & $0.0238^{+0.0038}_{-0.0031}\,(0.0263)$\\
$\Omega_c h^2$          & $0.1202^{+0.0083}_{-0.0065}\,(0.1225)$& $0.1262^{+0.0073}_{-0.0086}\,(0.129)$ \\
$\tau_{reio}$                  & $< 0.12 \,(0.10)$& $0.125\pm 0.055 \,(0.154)$ \\
 Age & $13.17^{+0.25}_{-0.37}\,(13.06)$&$13.14^{+0.26}_{-0.29}\,(13.10)$ \\
$ x_0$               & -& $-4.14^{+0.24}_{-0.18}\,(-3.93)$ \\
\hline
\hline
$ H_0 $               & $72.6^{+1.8}_{-1.3}\,(73.00)$& $72.9\pm 1.4\,(73.04)$ \\
$\sigma_8$               & $0.816^{+0.017}_{-0.021}\,(0.810)$& $0.813\pm 0.020\,(0.818)$ \\
$ S_8$               & $0.795\pm 0.017\,(0.801)$& $0.789\pm 0.018\,(0.7952)$ \\
$ \Omega_m$               & $0.285^{+0.011}_{-0.0083}\,(0.287)$& $0.2834\pm 0.0093\,(0.282)$ \\
\hline 
\hline
\textit{Planck} lensing & 8.81 &6.20\\
BAO &28.62 &  28.93\\
DES &  502.37   &   502.15\\
SN &  1034.83   &   1035.26\\
$H_0$ &  0.025   &   0.014\\
\hline
Total $\chi^2 $   & 1574.66 &  1572.56\\
$\Delta \chi^2 $ & - &  -2.1\\
\hline
\hline

\end{tabular}
\caption{ The mean $\pm 1 \sigma$(best-fit) constraints on the cosmological parameters inferred from the \texttt{Planck 2018 CMB lensing}, Baryon Acoustic Oscillations data from various surveys, Pantheon \texttt{SNIa} redshift-luminosity data, \texttt{Dark Energy Survey-Y1} and local measurements for $H_0$ by \texttt{SH0ES} team for $\Lambda$CDM, and UDE. Notice the higher value of $H_0$ and lower value of $S_8$.} \label{tab:Lensing+BAO+DES+SN+H0}
\end{table}

We find $H_0 = 70.22^{+0.49}_{-0.66}$, $S_8 = 0.810^{+0.011}_{-0.0080}$ and $\Omega_{m} = 0.2985^{+0.0054}_{-0.0044}$ for the UDE scenario. This brings down the $H_0$ and $S_8$ tensions to around $1.87 \sigma$ and $1.3 \sigma$ with $\texttt{SH0ES}$ and $\texttt{DES Y1}$ respectively for UDE model. For $\Lambda$CDM, we find $H_0 = 68.22 \pm 0.36$ , $S_8 = 0.812 \pm 0.0093$ and $\Omega_m = 0.3037 \pm 0.0047$ which are in $3.30 \sigma$ and $1.41 \sigma$ tension.
The $\chi^2$ statistics for each data set in this fit are presented in table \ref{tab:Base+Des}. The UDE improvement in the total $\chi^2$ -statistic is $\Delta \chi^{2} = +0.77$ which is not a better fit with one additional parameter, $x_0$.

\subsubsection{\texttt{Planck 2018 Lensing}+\texttt{BAO}+\texttt{DES}+\texttt{SN}+\texttt{H0}}

\begin{figure}[]
\centering
\includegraphics[scale=0.35]{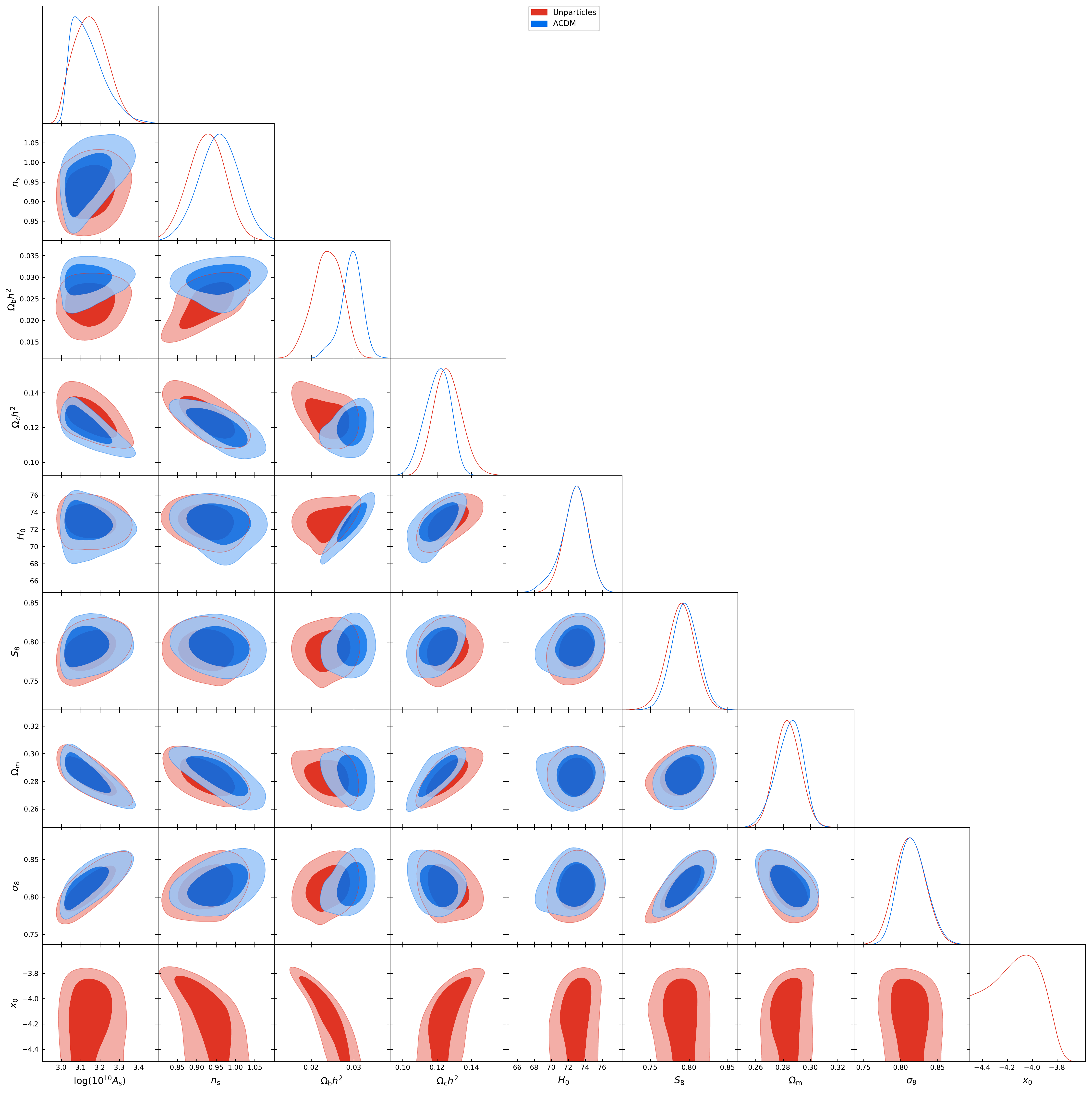}
\caption{Posterior distribution of cosmological parameters from \texttt{Lensing+BAO+DES+SN+$H_0$} data. We present the contour plots for $\Lambda$CDM and UDE in blue and red respectively with $1 \sigma$ and $2 \sigma$ CL.}
\label{fig:Lensing+BAO+DES+SN+H0}
\end{figure}

Next, we combine the data sets which are classified as low redshift (late time measurements), i.e. without Primary Planck 2018, which is driving both tensions. We point out that with this compilation, we get the highest value of the Hubble parameter, close to the SH0ES-only measurement. The posterior distributions are shown in Fig \ref{fig:Lensing+BAO+DES+SN+H0}, and parameter constraints are in Table \ref{tab:Lensing+BAO+DES+SN+H0}. The UDE model is favored by the data sets as there is $\Delta\,\chi^{2} = -2.1$ with respect to $\Lambda$CDM. For the UDE scenario, we find $H_0$ to be $72.9\pm 1.4$, which is  $1.63 $ $\sigma$ away from \texttt{Planck} only derived Hubble constant in Table \ref{tab:Planck 2018}. We find that $S_8 = 0.790 \pm 0.018$ which has almost $1.01 \sigma$ discrepancy from the \texttt{Planck 2018} value.

\subsubsection{\texttt{Primary \textit{Planck} 2018}+SN+$H_0$}
\begin{table}[!ht]
\textbf{ Constraints from \textit{Planck} 2018 CMB+SN+$H_0$}
   \vspace{1 em} \\
\centering
\begin{tabular}{|c|c|c|c|}

\hline
\hline

    Parameter         & $\Lambda$CDM & UDE  \\ 
\hline
\hline
$\ln(10^{10} A_\mathrm{s})$          & $3.047^{+0.016}_{-0.018}\,(3.043)$  &   $3.038\pm 0.018\,(3.034)$ \\
$n_\mathrm{s}$          & $0.9685^{+0.0056}_{-0.0050}\,(0.9723)$& $0.9725\pm 0.0053\,(0.9722)$ \\
$\Omega_b h^2$        & $0.02248^{+0.00021}_{-0.00017}\,(0.0225)$ &$0.02286\pm 0.00020\,(0.0229)$ \\
$\Omega_c h^2$          & $0.1187^{+0.0015}_{-0.0020}\,(0.117)$& $0.1251^{+0.0023}_{-0.0027}\,(0.126)$ \\
$\tau_{reio}$                  & $0.0566\pm 0.0086 \,(0.056)$& $0.0582\pm 0.0090\,(0.0557)$ \\
 Age & $13.775^{+0.028}_{-0.034}\,(13.75)$ & $13.31^{+0.14}_{-0.096}\,(13.30)$ \\
$ x_0$               & -& $-4.359^{+0.077}_{-0.092}\,(-4.34)$ \\
\hline
\hline
$ H_0 $               & $68.00^{+0.92}_{-0.70}\,(68.46)$& $71.29^{+0.96}_{-1.1}\,(71.18)$ \\
$\sigma_8$               & $0.8082^{+0.0079}_{-0.0088}\,(0.8045)$& $0.8132^{+0.0093}_{-0.0078}\,(0.813)$ \\
$ S_8$               & $0.817^{+0.017}_{-0.023}\,(0.8045)$& $0.803^{+0.018}_{-0.014}\,(0.806)$ \\
$ \Omega_m$               & $0.3069^{+0.0088}_{-0.013}\,(0.3005)$& $0.2926^{+0.0087}_{-0.0077}\,(0.295)$ \\
\hline 
\hline
low-$\ell$ TT & 22.06 &21.81\\
low-$\ell$ EE &396.23 & 396.15\\
high-$\ell$ TTTEEE &  2349.85   &   2350.61\\
SN &  1034.73   &   1034.77\\
$H_0$ &  13.25   &   2.41\\
\hline
Total $\chi^2 $   & 3816.14 & 3805.78\\
$\Delta \chi^2 $ & -  & -10.36\\ 
\hline
\hline
\end{tabular}
\caption{ The mean $\pm 1 \sigma$(best-fit) constraints on the cosmological parameters inferred from the \textit{Planck} 2018 CMB data ( \texttt{TTEETE}), Pantheon \texttt{SNIa} redshift-luminosity data and local measurements for $H_0$ by \texttt{SH0ES} team for $\Lambda$CDM and UDE.}
 \label{tab:Planck+SN+H0}
\end{table}
\begin{figure}[]
\centering
\includegraphics[scale=0.35]{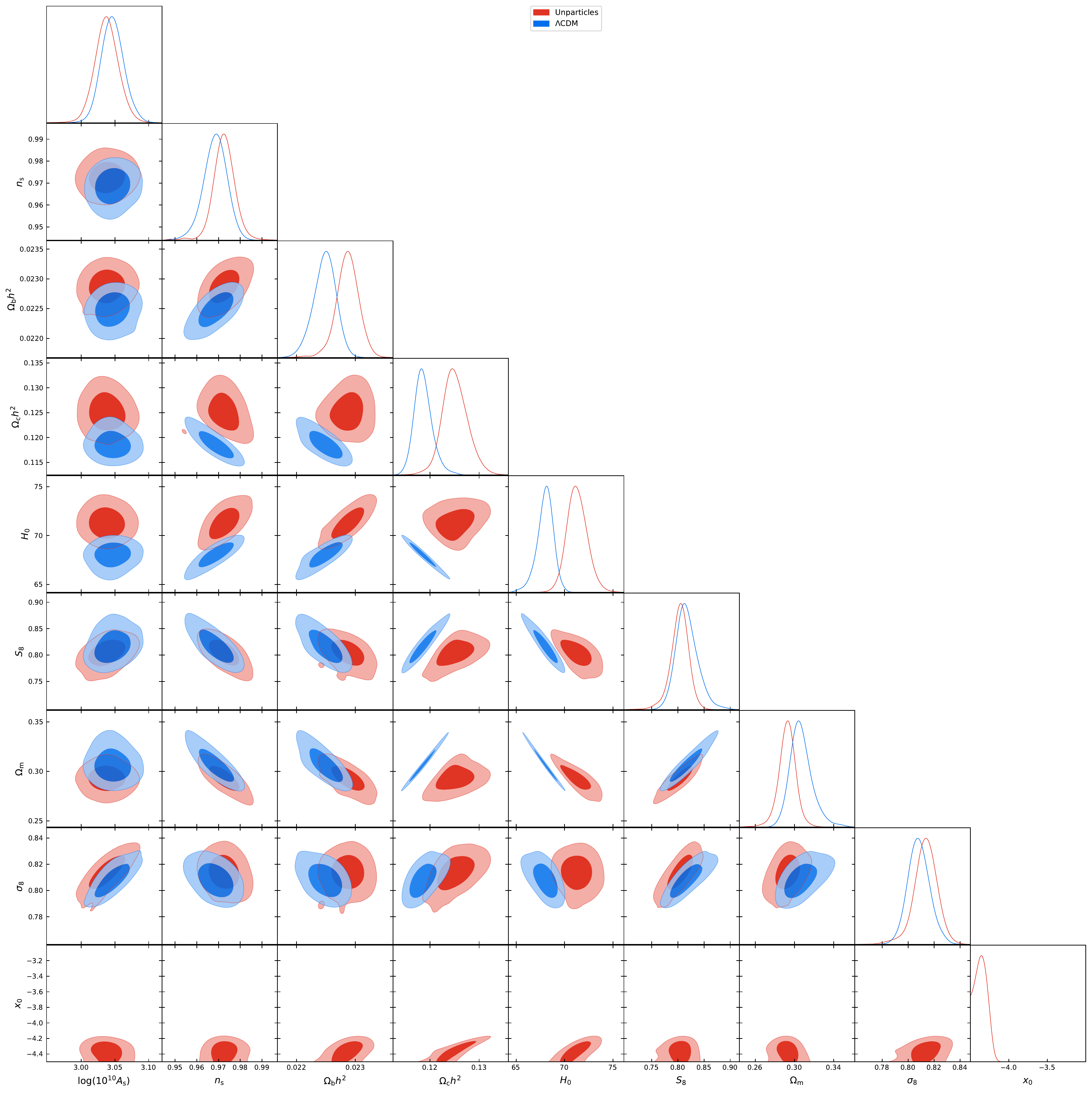}
\caption{Posterior distribution of cosmological parameters from the \texttt{Primary Planck 2018} + \texttt{SN} + \texttt{$H_0$}. We present the contour plots for $\Lambda$CDM and  UDE in blue and red respectively with $1 \sigma$ and $2 \sigma$ CL. Notice the notable difference in the preferred values of $H_0$ and $S_8$. }
\label{fig:Planck+SN+H0}
\end{figure}

We now jointly fit \texttt{Primary Planck 2018}, \texttt{SN}, and \texttt{$H_0$} as to avoid the inclusion of 
$f\sigma_8$ measurements in BAO measurements. We find $H_0=68^{+0.92}_{-0.70}$  and $71.29^{+0.96}_{-1.1}$ for $\Lambda$CDM and UDE respectively. For UDE $\Omega_m = 0.292^{+0.0082}_{-0.0077}$ , $S_8 = 0.803^{+0.018}_{-0.014}$ and  $\Omega_m = 0.3069^{+0.0088}_{-0.013} $ , $S_8 = 0.817^{+0.017}_{-0.023}$ for $\Lambda$CDM. The $H_0$ and $S_8$ tensions for UDE are $\simeq 0.95 \sigma$. 
The posterior distributions for this combination of data sets are shown in Fig \ref{fig:Planck+SN+H0}. The best-fit parameters and $ 68 \%$ CL constraints on cosmological parameters are tabulated in Table \ref{tab:Planck+SN+H0}.
The UDE scenario is most favored by this particular combination of data with $\Delta \chi^{\text{total}} = -10.36$. Again the improvement in the likelihood is mostly driven by the fit to the SH0ES data. This is the best improvement in the likelihood we have achieved and both tensions are reduced to less than one standard deviation.

\subsubsection{\texttt{\textit{Primary Planck} 2018}+\texttt{Lensing}+\texttt{BAO}+\texttt{SN}+\texttt{DES}+\texttt{$H_0$}}
\begin{table}[ht]
\textbf{ Constraints from \textit{Planck} 2018 CMB+\texttt{Lensing}+\texttt{BAO}+\texttt{SN}+\texttt{DES}+\texttt{$H_0$}}
   \vspace{1 em}  \\
\centering
\begin{tabular}{|c|c|c|c|}

\hline
\hline

    Parameter         & $\Lambda$CDM & UDE  \\ 
\hline
\hline
$\ln(10^{10} A_\mathrm{s})$          & $3.049\pm 0.016$\,(3.051) &    $3.040\pm 0.015$\,(3.037)\\
$n_\mathrm{s}$          & $0.9694^{+0.0050}_{-0.0037}$\,(0.971)&  $0.9708\pm 0.0039$\,(0.971)\\
$\Omega_b h^2$        & $0.02254^{+0.00020}_{-0.00013}$\,(0.02260) & $0.02283\pm 0.00017$\,(0.02284)\\
$\Omega_c h^2$          & $0.11800^{+0.00069}_{-0.0015} $\,(0.1176)&  $0.1249^{+0.0016}_{-0.0024}$\,(0.1249)\\
$\tau_{reio}$                  & $0.0586\pm 0.0086 $\,(0.060)& $0.0583\pm 0.0077$\,(0.0574) \\
 Age & $13.765^{+0.018}_{-0.032}$\,(13.75) & $13.35^{+0.12}_{-0.069}$\,(13.35)\\
$ x_0$               & -& $-4.380^{+0.038}_{-0.12}$\,(-4.37) \\
\hline
\hline
$ H_0 $               & $68.28^{+0.71}_{-0.31}$\,(68.47)&  $70.87^{+0.61}_{-0.79}$\,(70.76)\\
$ \sigma_8 $               & $0.8071\pm 0.0059$\,(0.8069)&  $0.8142\pm 0.0073$\,(0.8140)\\
$ S_8$               & $0.8111^{+0.0080}_{-0.015}$\,(0.8075)&  $0.808^{+0.010}_{-0.0078}$\,(0.8093)\\
$ \Omega_m$               & $0.3030^{+0.0038}_{-0.0091}$\,(0.3004)&  $0.2956^{+0.0049}_{-0.0041}$\,(0.2965)\\

\hline
\hline
\textit{Planck} & 2777.28 &2777.73\\
BAO &29.299 & 29.97\\
DES &  508.22   & 509.06\\
SN &  1034.74    & 1034.74\\
$H_0$ &  13.20   & 3.52\\
\hline
Total $\chi^2 $   & 4362.75 & 4355.03\\
$\Delta \chi^2 $ & - & -7.72\\
\hline
\hline
\end{tabular}
 
\caption{ The mean $\pm 1 \sigma$(best-fit) constraints on the cosmological parameters inferred from the \textit{Planck} 2018 \texttt{TTEETE} , \texttt{Planck 2018 CMB lensing}, BAO, Pantheon, \texttt{Dark Energy Survey -Y1} and $H_0$ by \texttt{SH0ES} team for $\Lambda$CDM and UDE.} \label{tab:Base+Des+H0}
\end{table}

\begin{figure}[]
\centering
\includegraphics[scale=0.35]{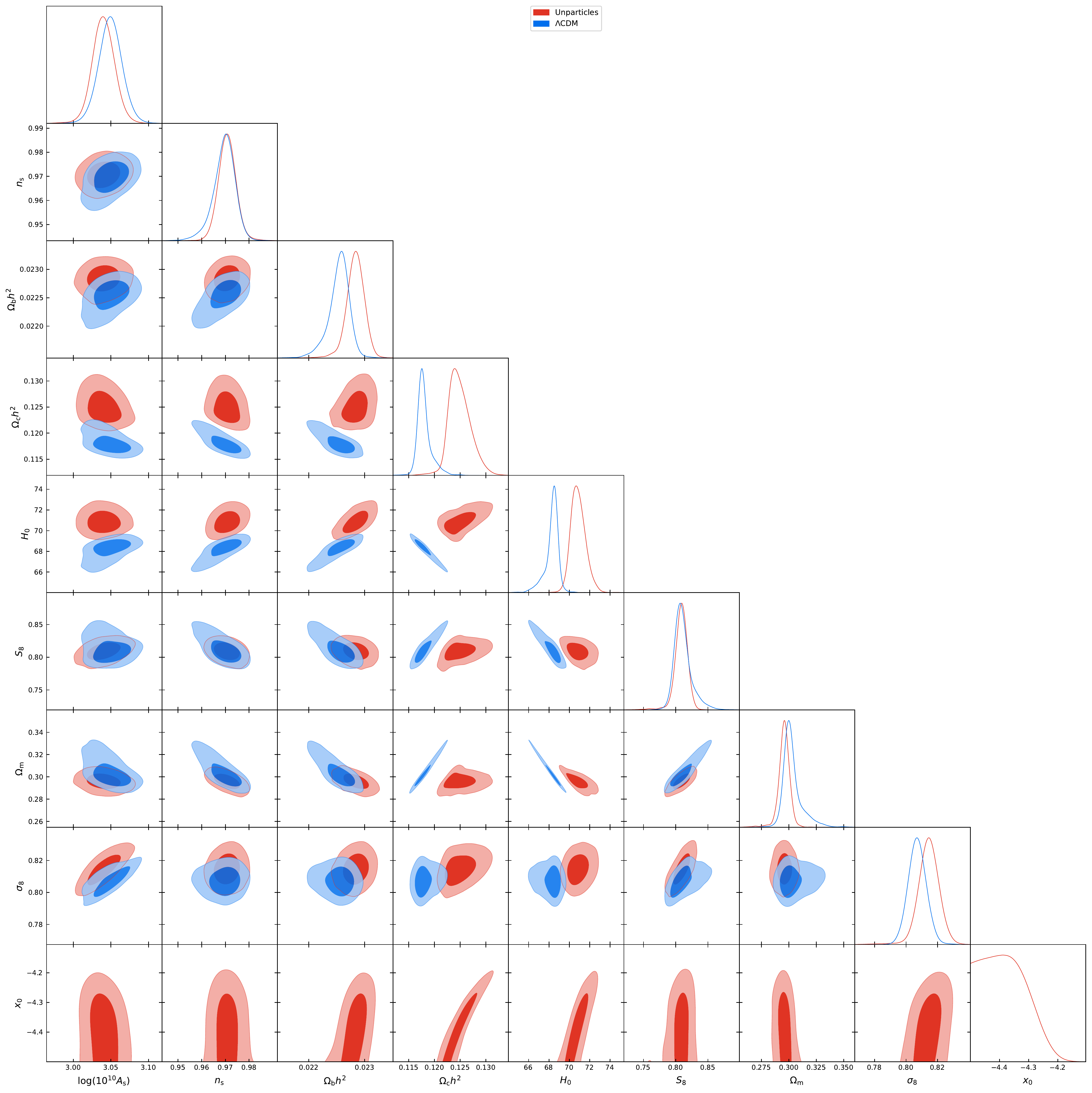}
\caption{Posterior distribution of cosmological parameters from the \textit{Planck} 2018 primary CMB+ Lensing+BAO+SN+DES+$H_0$. We present contour plots for $\Lambda$CDM and UDE in red and blue respectively with $1 \sigma$ and $2 \sigma$ confidence contours.}
\label{fig:Planck+SN+BAO+DES+H0}
\end{figure}

Finally, we combine all data - Primary \textit{Planck} 2018, \textit{Planck} 2018 CMB lensing, BAO, DES, Pantheon, and SH0ES data. The Posterior distributions for this combination of data sets are shown in Fig \ref{fig:Planck+SN+BAO+DES+H0}. The best-fit parameters and $ 68 \%$ CL constraints on cosmological parameters are tabulated in Table \ref{tab:Base+Des+H0}.  We find $H_0=70.87^{+0.61}_{-0.79}$ compared to $H_0= 68.28^{+0.71}_{-0.31}$ in
$\Lambda$CDM and 
$S_8=0.808^{+0.010}_{-0.0078}$ compared to $0.8111^{+0.0080}_{-0.015}$ for $\Lambda$CDM. The resulting tensions in $H_0$ and $S_8$ with respect to \texttt{SHOES} and \texttt{DES-Y1} are 1.66 $\sigma$ and 1.25$\sigma$ respectively. The best-fit value for the unparticles parameter $x_0=-4.380^{+0.038}_{-0.12}$ and there is a significant improvement in the likelihood of $\Delta \chi^2=-7.72$.

\begin{figure}[]
\centering
\includegraphics[width=12cm]{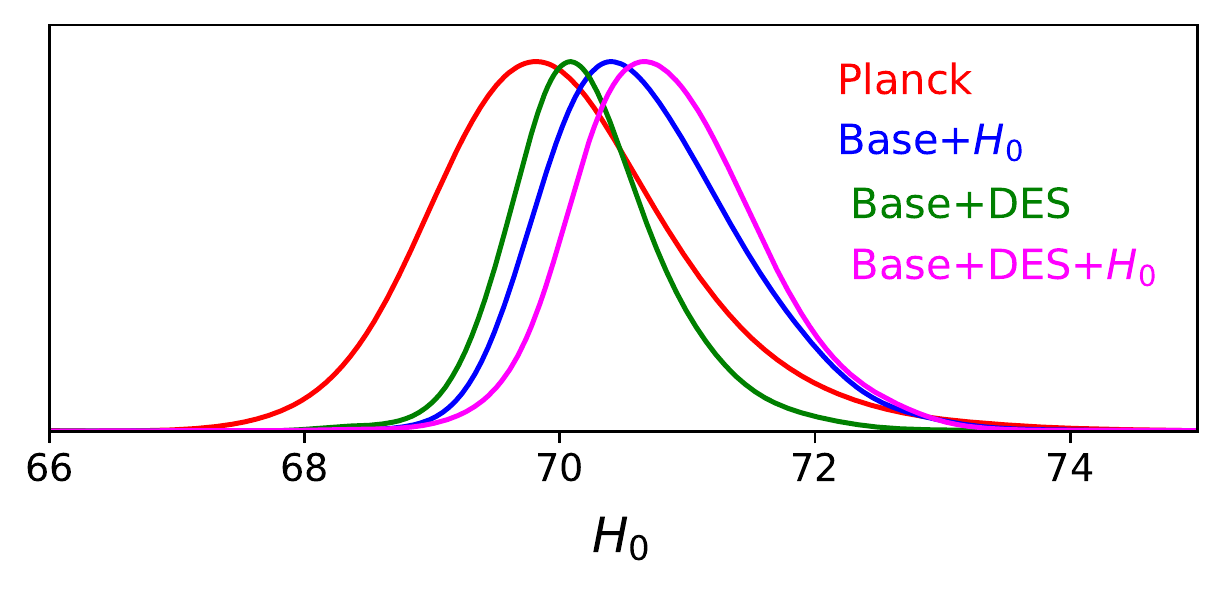}
\includegraphics[width=12cm]{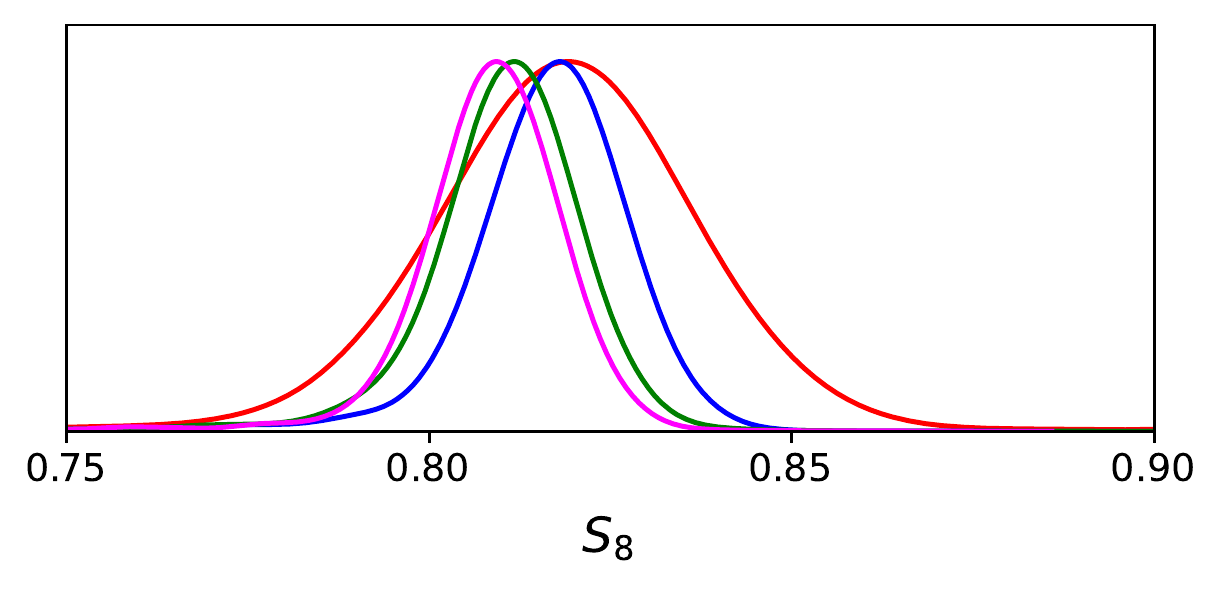}
\includegraphics[width=12cm]{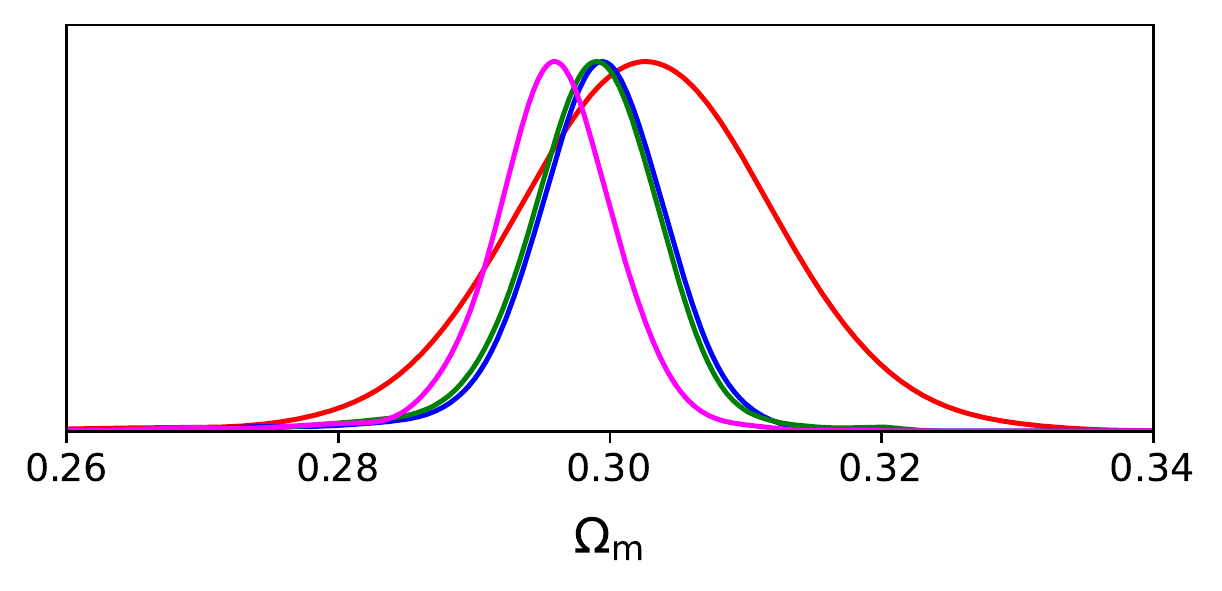}
\caption{Comparison of posteriors of $H_0$ (upper), $S_8$ (middle), and $\Omega_m$ inferred from the different data-sets for the UDE scenario. The caption "Base" means Primary \textit{Planck} 2018+Lensing+BAO+SN data. }
\label{fig:camparison1}
\end{figure}
\begin{figure}[]
\centering
\includegraphics[width=8cm]{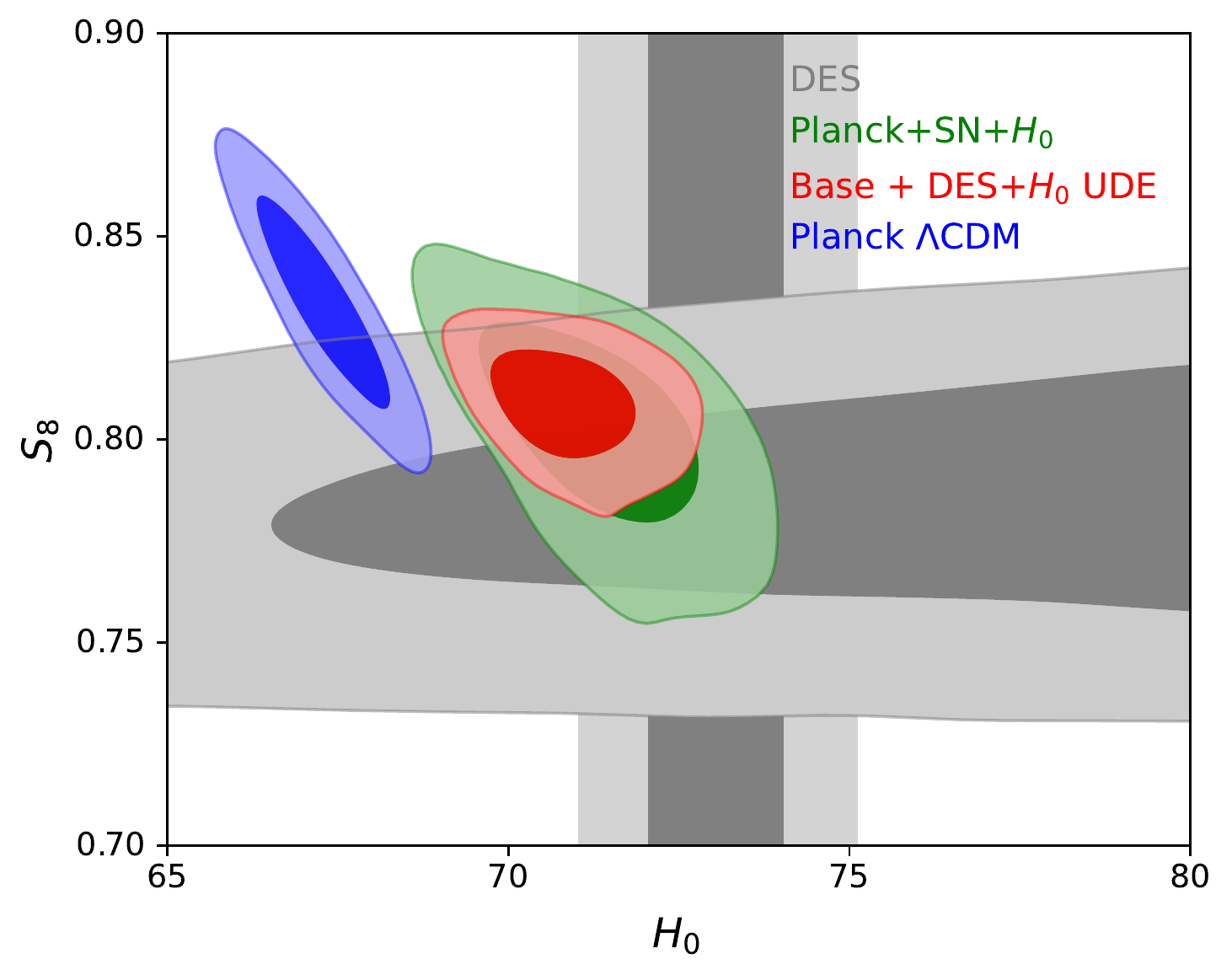}
\includegraphics[width=8cm]{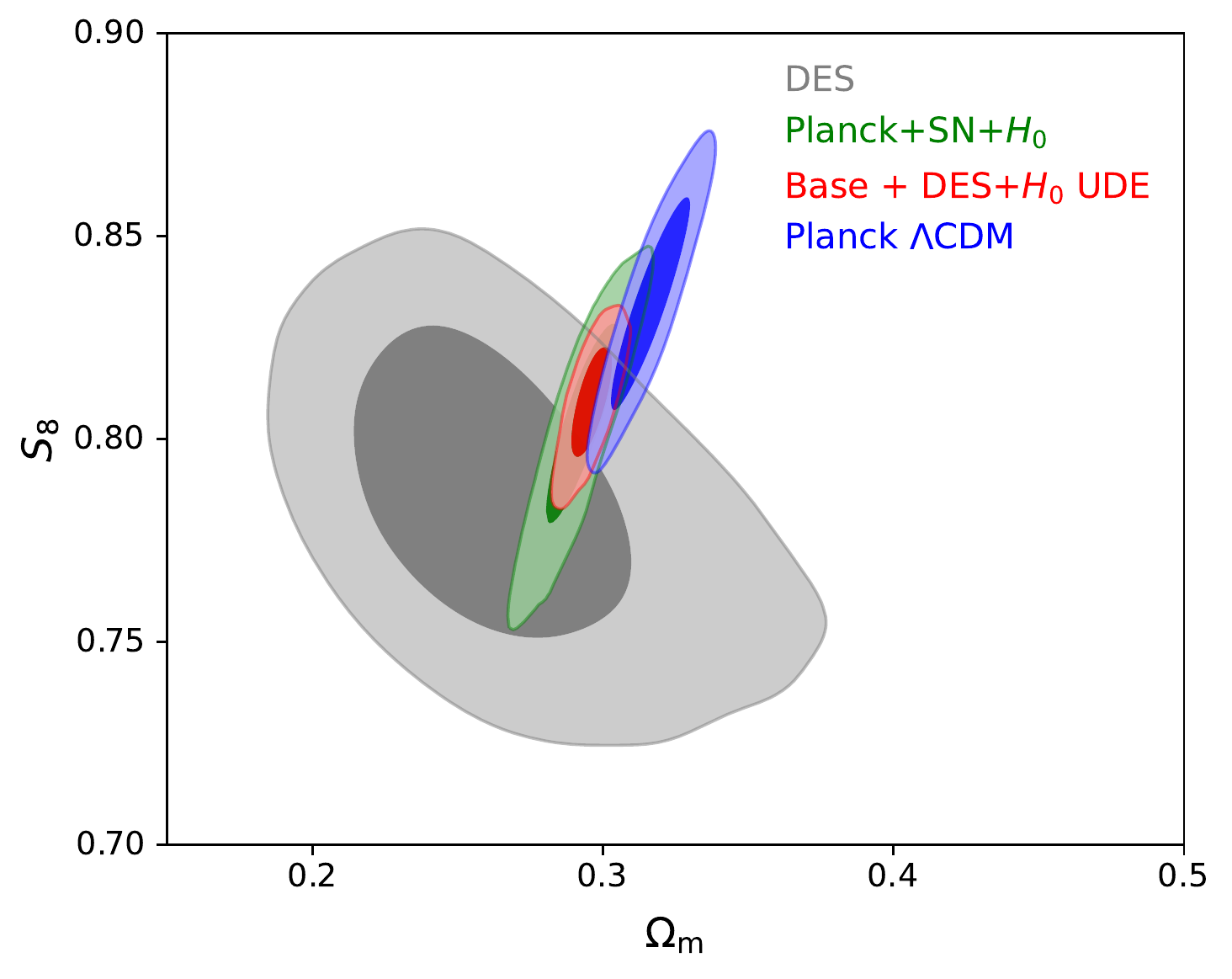}
\caption{The 2-D marginalized PDF contours for $H_0 - S_8$ and $\Omega_m - S_8$ in left and right panel respectively. We compare the alleviated $H_0$ and $S_8$ tensions in the UDE scenario with respect to \texttt{SHOES} and \texttt{DES} within $\Lambda$CDM. The \texttt{SHOES} and \texttt{DES} measurements are represented in Gray. The Unparticles results are shown in green and red for CMB Planck 2018 and All data sets respectively while Planck $\Lambda$CDM is denoted in blue. In fig, base means the combination of CMB Planck 2018 + BAO + SN measurements.}
\label{fig:camparison2}
\end{figure}

\section{Conclusions}
\label{sec:conclusions}
We have discussed the possibility of unparticles as viable Dark Energy candidate, in the UDE model. As the Banks-Zaks theory is displaced from its conformal fixed point, the energy density and pressure of the fluid have two terms resulting in a temperature-dependent equation of state. As a result, along with the dynamics of the Universe, the fluid has a limiting temperature, which causes unparticles to evolve as a radiation-like fluid and asymptote to a CC in early and late times, respectively. The consideration of unparticles as a DE source makes the theory immune to cosmic coincidence and no-dS conjecture. A major point is the fact that the DE behavior is emergent - due to the thermodynamical behavior of the fluid, and not because of some specific degree of freedom in vacuum or a modification of gravity.  

We further investigated the challenges posed by Hubble ($H_0$) and large-scale structure $S_8$ tension in the era of precision cosmology. In this work, we analyzed the suggestion that the UDE scenario accounts for the discrepancy between various data sets.
We find that the UDE scenario can alleviate both tensions. We also find evidence of obtaining the Unparticles in the Universe, such that $x_0\simeq -4.3$, or more conservatively $-5<x_0<-4$. 

To quantify the evidence of UDE, we have shown the effect of UDE parameter $x_0$ on the CMB temperature anisotropy and matter power spectra in Figures \ref{fig:cll} and \ref{fig:pk}. 
The change in power spectra of both CMB temperature anisotropy and matter drives the change of values of $H_0$ and $S_8$. From the unparticles' temperature today we can derive the amount of extra radiation added in the early universe due to unparticles \cite{Artymowski:2021fkw}, with $\Delta N_{eff}=0.216$ at decoupling.
We plot the resulting 1D distribution for the relevant parameters $H_0, S_8$ and $\Omega_m$ in figure \ref{fig:camparison1}. All combinations show a higher $H_0$ and lower $S_8$ than the Planck result for $\Lambda$CDM reducing both tensions. 
With the exception of Primary Planck data alone (and to a much lesser extent the combination of data sets excluding the SH0ES result), the UDE not only shows a significant reduction in both tensions, but a considerably better fit to the data with $\Delta \chi^2=-2.1$$\textbf{--}$$\,-10.4$. Specifically, combining all data sets gives $H_0=70.87^{+0.61}_{-0.79}$, $S_8=0.808^{+0.01}_{-0.0078}$, and $x_0\simeq -4.38$, while the best improvement of likelihood came from the subset of Planck+Pantheon+SH0ES with $H_0=71.29^{+0.96}_{-1.1}$, $S_8 = 0.803^{+0.018}_{-0.014}$ and $x_0\simeq-4.36$.
We further demonstrate the reduction of the tension in figure \ref{fig:camparison2}. The figure shows $\Lambda$CDM Planck 2018 results in blue, the SH0ES and DES data in gray, and our results for Planck+Pantheon+SH0ES and for all data in green and red respectively by plotting 2D likelihoods. In the left panel, $S_8$ and $H_0$ are shown and $S_8$ and $\Omega_m$ are on the right panel. To summarize, the model is restoring the cosmological concordance.

It would be interesting to further test the UDE model. This can proceed in several ways.  First, since the UDE behaves as radiation at early times, at a certain point such extra radiation should appear in the form of $\Delta N_{eff.}$, which should be measured (consistent with the value of $x_0$). Second, going to higher redshifts in galaxy surveys and weak lensing experiments up to $z\sim 5$ will allow us to measure the time dependence of the equation of state parameter $w(z)$ thus confirming or ruling out the model. These measurements will also improve the measurement of the growth factor $f$ where there is also some difference between $\Lambda$CDM and UDE \cite{Artymowski:2020zwy}. Third,  considering the effect of UDE on ISW at early and late times \cite{Krolewski:2021znk,Vagnozzi:2021gjh,Cabass:2015xfa}. 
Fourth, evaluating carefully the effects suppressed couplings between the Banks-Zaks theory and the Standard Model may have. A first interesting attempt has been carried out in \cite{vanPutten:2022may}. Fifth, going beyond the fluid approximation one may consider microscopic effects that can be observed. Sixth, one would like better theoretical control of the model, where $\delta$ can be derived, rather than being a free parameter.

The "problem" of UDE until now has been that the deviations may be so small compared to $\Lambda$CDM that they will not be detectable.
The present work shows that at least part of the parameter space of the model, with $-5<x_0<-4$ is favorable since it considerably reduces existing cosmological tensions and is a statistically significant better fit to the data.
Going beyond the specific model, we can try and learn some general lessons for model building for treating the Hubble and LSS tension. It seems an important feature is the change in the equation of state $w(z)$ from radiation to DE. It would be interesting to consider other models that exhibit such behavior and better yet analyzing such a phenomenon in a model-independent way which we plan to do next.

\section*{Acknowledgements}
We acknowledge the Ariel HPC Center at Ariel University for providing computing resources that have contributed to the research results reports reported within this paper.
\bibliographystyle{JHEP}
\bibliography{reference}

\end{document}